\documentclass[showpacs, prb,twocolumn,preprintnumbers , superscriptaddress, aps]{revtex4-2}
 
\usepackage{color}
\usepackage{amsmath,amssymb}
\usepackage{pifont}
\usepackage{amssymb}  
\usepackage{bbold}
\usepackage{float}
\usepackage[caption=false]{subfig}
\usepackage{amsmath,amssymb}
\usepackage{mathtools}
\usepackage{physics}
\usepackage{breqn} 

\usepackage[caption=false]{subfig}
\usepackage{tikz}
\usepackage{makecell}
\usepackage{pifont}   
\usepackage{graphicx} 
\usepackage{dcolumn}  
\usepackage{bm}       
\usepackage{multirow} 
\usepackage{placeins}
\usepackage[colorlinks]{hyperref}
\usepackage{mathtools}

\usepackage{breqn}
\usepackage{mathrsfs}

\usepackage{multirow}

\begin{document}

\title{Fingerprints of an Imbert-Fedorov-like effect in the tunneling transmission of Rarita-Schwinger semi-metals}

    \author{Otman Bouladiane}
    \affiliation{Laboratory of Theoretical Physics$,$ Faculty of Sciences$,$ Choua\"ib Doukkali University$,$ PO Box 20$,$ 24000 El Jadida$,$ Morocco}
  
    \author{Ahmed Jellal}
  \affiliation{Laboratory of Theoretical Physics$,$ Faculty of Sciences$,$ Choua\"ib Doukkali University$,$ PO Box 20$,$ 24000 El Jadida$,$ Morocco}
        \author{Hocine Bahlouli}
  \affiliation{Physics Department and IRC Advanced Materials$,$
  King Fahd University
  of Petroleum $\&$ Minerals$,$
  Dhahran 31261$,$ Saudi Arabia}
  \author{A. Al Luhaibi}
        \affiliation{Physics Department$,$
		King Fahd University
		of Petroleum $\&$ Minerals$,$
		Dhahran 31261$,$ Saudi Arabia}
        \affiliation{Interdisciplinary Research Center (IRC) for Advanced Quantum Computing (AQC)$,$ KFUPM$,$ Dhahran$,$ Saudi Arabia}
    \author{Michael Vogl}
  \affiliation{Physics Department$,$
  King Fahd University
  of Petroleum $\&$ Minerals$,$
  Dhahran 31261$,$ Saudi Arabia}
      \affiliation{Interdisciplinary Research Center (IRC) for Advanced Quantum Computing (AQC)$,$ KFUPM$,$ Dhahran$,$ Saudi Arabia}

\begin{abstract}
We study tunneling through a square barrier in a rotationally symmetric Rarita-Schwinger semi-metal and identify fingerprints of an Imbert-Fedorov-like effect in the tunneling transmission. The problem is intrinsically multichannel at all energies because there exist two propagating sectors with spin projections $m=1/2$ and $m=3/2$. We derive the transmission amplitudes analytically and compare the single-channel and coherent mixed-incidence cases in the two channels. Interestingly, we observe that tunneling at high energy shows a bias towards scattering into spin projection $3/2$ contributions. For single-channel injection, we find that the transmission remains symmetric under a mirror transformation of the incident angle. In contrast, for the coherent superposition, we find a directional asymmetry in the transmission probability $T(k_y)\neq T(-k_y)$. Importantly, this effect does not originate from an anisotropy of the dispersion. Instead, it arises from the phase structure of the multicomponent scattering states. The two spin projection sectors exhibit different scattering and barrier-propagation phases, which enter as interference terms when both channels appear as a coherent superposition. This interference term is identified as the cause of the broken mirror symmetry. We therefore discover an analog of the Imbert-Fedorov effect, at the level of interface-induced phase differences between internal wave components. Our result demonstrates that asymmetric tunneling can already occur in transport experiments, even in an idealized, isotropic multiband system, and should therefore not be automatically attributed solely to explicit band anisotropy.
\end{abstract}

\maketitle

\section{Introduction}

Studies of tunneling phenomena \cite{RevModPhys.80.1337,PhysRevLett.111.066803} have been fundamental to quantum mechanics since its inception. They are important because they directly demonstrate the wave nature of matter and the failure of classical intuition. Such effects are central to understanding quantum coherence, phase accumulation, and interference \cite{PhysRevB.68.205423,Dyakov2024ChiralLight}. A good understanding of tunneling is essential, since it appears in many physical systems, including atomic, molecular, and condensed-matter systems. Additionally, tunneling can offer insight into deeper quantum effects, where internal degrees of freedom, symmetry, and topology \cite{Culcer2020Transport2D,Chang2018ChiralTopological} can have significant influence.

The discovery of graphene \cite{doi:10.1126/science.1102896,Geim2007} was instrumental in reigniting interest in tunneling phenomena because analysis of tunneling coefficients revealed counterintuitive features absent in conventional Schr\"odinger systems. It also motivated broader investigations of tunneling in semi-metals exhibiting similar behavior. Such tunneling features are commonly attributed to two main ingredients i) the linear energy dispersion and ii) the chiral nature of Dirac-Weyl quasiparticles \cite{Katsnelson2006,RevModPhys.81.109}. More generally, similar behavior is found in so-called Dirac materials whose low-energy excitations are described by the massless Dirac equation and behave like relativistic particles. Prominent examples include graphene, topological insulators, and Weyl semi-metals. The exotic phenomena exhibited by these materials include Klein tunneling \cite{DOMBEY199941,Pereira_2010} (perfect transmission at normal incidence), strongly angle-dependent transmission, and, in graphene $pn$ junctions, negative refraction \cite{Katsnelson2006,Young2009}. All of these effects demonstrate that, in Dirac materials, effectively massless charge carriers behave very differently from conventional massive electrons when tunneling through potential barriers. More importantly, these results show that tunneling in semi-metals is not determined solely by barrier properties, such as geometry, but also by quasiparticle properties, including chirality, internal quantum numbers, and phase structure. This observation suggests that it is fruitful to explore new transport regimes based on symmetry, topology, and relativistic quantum dynamics \cite{Katsnelson2006ZB,RevModPhys.81.109}.

Beyond graphene, quantum tunneling has been studied in other semi-metals, including Dirac, Weyl, multi-Weyl, and nodal-line, and anisotropic semimetals \cite{RevModPhys.90.015001,PhysRevResearch.2.013088,BOUHLAL2021168563,PhysRevB.89.075124,Bouhlal_2022,PhysRevB.92.081201}. In these systems, results depend crucially on the band structure, topology, and internal degrees of freedom \cite{RevModPhys.90.015001}. In Dirac and Weyl semi-metals, transport is strongly shaped by chirality and topology \cite{PhysRevResearch.2.013088,PhysRevB.89.075124}, and effects related to Klein tunneling also appear. In nodal-line semi-metals, tunneling depends on how the dispersion varies around the nodal line and on which parts of the nodal ring are selected by transverse momentum \cite{PhysRevB.92.081201}. As a result, transmission can show strong momentum and angle selectivity. Higher-spin or otherwise complex semi-metals are especially interesting since multiple propagation channels may coexist at the same energy \cite{MANDAL2020126666,Mandal2023,PhysRevB.101.184503}. This allows interference and mode conversion between channels, while wave effects such as Fabry-Pérot resonances remain important \cite{Katsnelson2006}. Thus, tunneling in semi-metals involves not only conventional wave interference but also new phenomena associated with internal degrees of freedom.

To better understand internal-state effects intuitively, we can compare electronic transport with optical wave phenomena. Optical reflection, refraction, and beam propagation are clear examples of how internal wave structure affects scattering and they play a role in our everyday experience. Electronic tunneling is known to often mirror optical phenomena such as total internal reflection and Goos-Hänchen shifts \cite{PhysRevLett.102.146804,PhysRevB.62.10696,RAZA2025130167} - further motivating comparison between optical and electronic effects. Beyond scalar wave effects, multicomponent waves also matter. In optics, polarization plays this role, whereas in electronic systems, it is played by spin, pseudospin, or general internal channel structure \cite{PhysRevLett.96.073903,PhysRevLett.93.083901}. These analogies, we stress, are useful since they offer an intuitive way to discuss interface phenomena in multiband semi-metals.
They also motivate our search for similar effects in multiband semi-metals. Specifically, we are interested in transport effects arising from the interplay between multiple propagation channels and the internal wave-function structure. This interplay can produce unconventional transport features beyond those expected from band-structure symmetry alone. 

In the present work, we therefore investigate high-spin Rarita-Schwinger quasiparticles.
The optical phenomenon that serves as our main point of comparison is the Imbert-Fedorov effect \cite{PhysRevD.5.787,Fedorov_2013}. In optics, this effect refers to a small transverse shift of a finite beam upon reflection or refraction at an interface, due to different polarization components acquiring different scattering phases. We stress that we do not directly evaluate
such a real-space wavepacket shift. Rather, we use the
term “Imbert-Fedorov-like” to describe an analogous phase sensitive
mechanism: the two propagating spin projection sectors (later often abbreviated to just sectors) of the Rarita-Schwinger problem are transformed differently by the barrier, acquiring distinct scattering phases. When only one sector is incident, this has no consequence beyond an overall phase. For a coherent superposition of sectors, however, the relative phase becomes observable through interference and leaves a directional imprint on the tunneling probability. Thus, the observable studied
here is an angular, or equivalently transverse-momentum,
asymmetry of plane-wave transmission rather than a realspace
coordinate shift. Importantly, this effect appears even though both the barrier and the low-energy dispersion are symmetric. Taken together with coherent mixing between the two spin sectors, this leads to asymmetric tunneling probabilities.

The remainder of this paper is organized as follows. In Section \ref{Model}, we introduce the model and the tunneling setup. In Section \ref{results}, we present the analytical treatment and the main results for the transmission probability. In Section \ref{Explanation}, we explain the origin of the resulting asymmetry and discuss its relation to the Imbert-Fedorov effect. In Section \ref{sec:higherspin_bias}, we observe an interesting bias of scattering toward spin projection $3/2$ components. Finally, in Section \ref{Conclusion}, we summarize our conclusions.

\section{Model and setup}
\label{Model}

Our starting point is the so-called Rarita-Schwinger Hamiltonian, which describes effective spin-$\frac32$ excitations and originally appeared in the context of high-energy physics. It is given by \cite{PhysRevLett.124.127602,PhysRevB.101.184503,Mandal_2024,PhysRevB.105.235403}
\begin{equation}
    H(\bm{p}) = \sum_i p_i \left(v_1 J_i + v_2 J_i^3\right),
    \label{eq:eq1}
\end{equation}
where $\bm p$ denotes crystal momentum measured from the band-crossing point, $v_1$ and $v_2$ are velocity parameters, and $J_i$ are spin-$\frac32$ matrices. We stress that our work focuses on material realizations whose low-energy physics near a single fourfold crossing is described by Eq.~\eqref{eq:eq1}, without additional fixed-vector or multivalley couplings. Throughout this work, we use units with $\hbar=1$.

A convenient tensor-product representation of the spin-$\frac32$ matrices is
\begin{equation}
    J_z=\sigma_z\otimes \mathbb{1}+\frac12\,\mathbb{1}\otimes\sigma_z,
\end{equation}
together with
\begin{equation}
    J_\pm=J_x\pm iJ_y
    =\sqrt{3}\,\mathbb{1}\otimes\sigma_\pm
    +2\,\sigma_\pm\otimes\sigma_\mp,
\end{equation}
 where $\sigma_i$ are Pauli matrices, $\sigma_\pm=(\sigma_x\pm i\sigma_y)/2$, and the tensor products are used only as a compact representation of the spin-$\frac32$ algebra rather than to denote distinct physical degrees of freedom.

Density-functional-theory calculations reveal several material candidates that may host this kind of exotic massless fermion, with examples appearing in the family of transition-metal silicides such as CoSi and RhSi, as well as in PdBiSe \cite{PhysRevB.99.241104,PhysRevLett.119.206402}. This is significant because it opens the possibility of studying Rarita-Schwinger quasiparticles experimentally in condensed-matter systems, even though no fundamental Rarita-Schwinger particle is realized in the Standard Model. Rarita-Schwinger semi-metals are just one example of the more general observation that condensed-matter systems frequently provide effective low-energy realizations of quasiparticles that have no direct counterpart among elementary particles.

In all that follows, we restrict ourselves to the rotationally symmetric case $v_2=0$. The reasons for this restriction are threefold. First, it permits a fully analytical treatment. Second, the phenomenology is already rich in this limit, while including $v_2\neq0$ would substantially broaden the scope and may warrant a separate study. Third, the $v_2=0$ limit establishes a rotationally symmetric reference problem by removing band anisotropy. As we will see below, this allows us to distinguish effects arising from the internal degrees of freedom from those that could be caused by band anisotropy.

The barrier considered below depends only on $x$, and hence both components $k_y$ and $k_z$ of the momentum parallel to the interface are conserved. Their conservation alone does not allow either component to be neglected. However, in the rotationally symmetric case $v_2=0$, the Hamiltonian and the $x$-normal barrier are invariant under rotations about the $x$ axis. By simultaneously rotating the transverse momentum and the spin-$3/2$ state, an arbitrary $(k_y,k_z)$ can therefore be mapped to $(k_\perp,0)$, where $k_\perp=\sqrt{k_y^2+k_z^2}$. We may thus choose the $x$-$y$ plane as the scattering plane and set $k_z=0$ without excluding modes with nonzero transverse momentum. In this coordinate frame, the Hamiltonian reduces to \cite{PhysRevB.101.184503,MANDAL2020126666,doi:10.1126/science.aaf5037}
\begin{align}
H =
\begin{pmatrix}
0 & \frac{\sqrt{3}v_1}{2}\pi_- & 0 & 0 \\
\frac{\sqrt{3}v_1}{2}\pi_+ & 0 & v_1\pi_- & 0 \\
0 & v_1\pi_+ & 0 & \frac{\sqrt{3}v_1}{2}\pi_- \\
0 & 0 & \frac{\sqrt{3}v_1}{2}\pi_+ & 0
\end{pmatrix},
\label{ham1}
\end{align}
where $\pi_\pm=p_x\pm ip_y$.

\begin{figure}[htbp!]
\centering
\includegraphics[width=0.8\columnwidth]{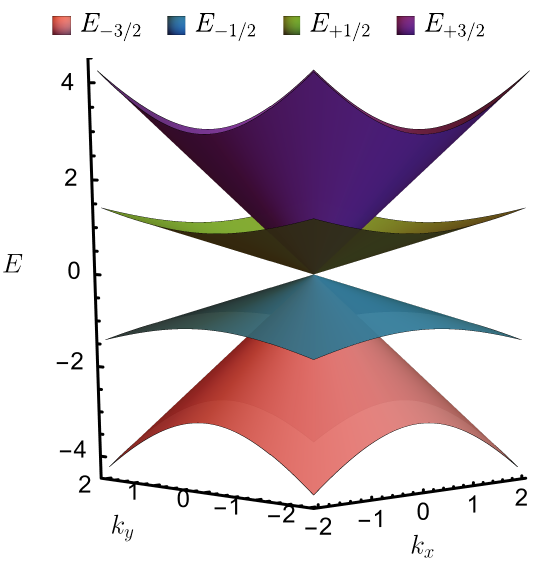}
\caption{Energy dispersion of the spin-$3/2$ Rarita–Schwinger semi-metals as a function of momentum ($k_x,k_y$). The four bands $E_{\pm 3/2}$ and $E_{\pm 1/2}$ form Dirac-like cones centered at $k = 0$, with distinct slopes reflecting their different group velocities.}
\label{Enerdisp}
\end{figure}

Excitations of Rarita-Schwinger semi-metals are particularly interesting because they lead to two different projections of the effective spin, namely $\pm\frac32$ and $\pm\frac12$ \cite{PhysRevB.93.045113,PhysRevLett.119.206402}. As a result, the spectrum exhibits four linearly dispersing modes, which may be viewed as two Dirac-like cones intersecting at a single point in momentum space. Importantly, these cones have different slopes, as shown in Fig.~\ref{Enerdisp} with a 3D plot of the energy dispersion as a function of momentum ($k_x,k_y$). The two cones 
correspond to two characteristic velocities, $v_1/2$ and $3v_1/2$. The steeper branch is associated with spin projection $\frac32$, while the flatter branch is associated with spin projection $\frac12$.

\begin{figure}[ht]
\centering
\includegraphics[width=\columnwidth]{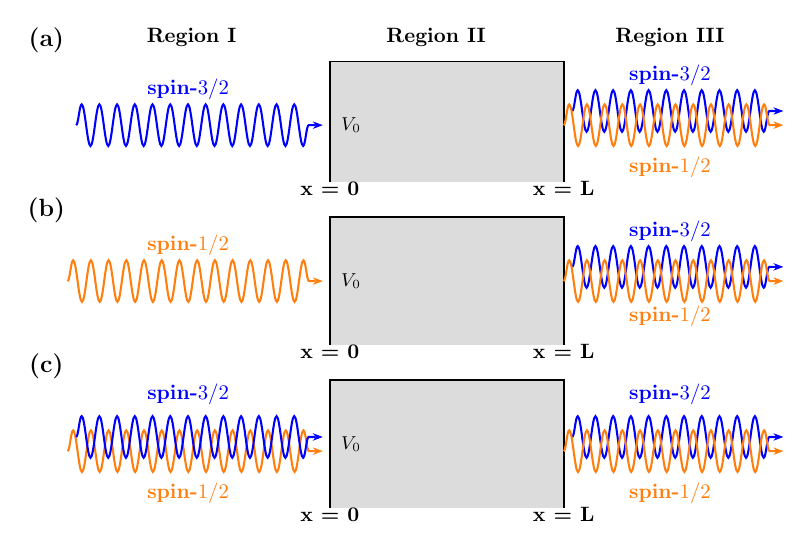}
\caption{Tunneling setup for three incident configurations:
(a) a single incoming channel with spin projection $m=3/2$,
(b) a single incoming channel with spin projection $m=1/2$,
and (c) a coherent superposition of both channels.
The Rarita--Schwinger fermion encounters a square barrier of
height $V_0$ and width $L$. Both transmitted branches are shown
because the barrier can induce interbranch conversion.
Their amplitudes and relative phase depend on the incident
configuration; identical output arrows do not imply identical
transmitted states.}
\label{Tunprob}
\end{figure}

In this work, we consider tunneling through a square potential barrier of height $V_0$ and width $L$,
\begin{align}
V(x)=
\begin{cases}
V_0, & 0<x<L,\\
0, & \text{otherwise}.
\end{cases}
\end{align}
Here, $V(x)$ represents an idealized effective barrier assumed to be
uniform parallel to its interfaces. This approximation applies when
transverse variations of the potential are negligible over the active
region.
Although tunneling in spin-$\frac32$ systems such as Rarita-Schwinger semi-metals has already been studied, most previous transport works considered incident waves from different spin-projection sectors separately, rather than coherent mixtures of them \cite{MANDAL2020126666,Mandal2023}. We will see that this restriction misses much of the interesting phenomenology that we discuss here. Specifically, for the double-cone band structure considered here, it is natural to ask what happens when both propagating sectors are present already at low energy and are injected coherently. As we will show, this gives rise to qualitatively new effects. We will therefore consider and compare three tunneling setups shown schematically in Fig.~\ref{Tunprob}.
In the next section, we solve the tunneling problem in all three cases.

\section{Observations and results}
\label{results}
\subsection{Analytical solution}
In Rarita-Schwinger semi-metals, states at a given energy generally involve two propagating sectors with spin projections $m=1/2$ and $m=3/2$. A barrier breaks translation invariance along $x$, so $p_x$ is no longer a good quantum number. Scattering can mix the two sectors. By contrast, translation invariance along $y$ remains. Therefore, the transverse momentum $k_y$ is conserved throughout the tunneling process (we use letters $k_i$ for the eigenvalues of operators $p_i$). In other words, we may express wavefunctions as
\begin{equation}
\Psi(x,y)=e^{ik_y y}\,\widetilde\Psi(x).
\end{equation}
This observation reduces the problem to one-dimensional scattering for the four-component spinor $\widetilde\Psi(x)$. Throughout this subsection, we stress, like before, we use units with $\hbar=1$.

Rather than solving the coupled equations component by component, it is more transparent to work directly with the bulk eigenspinors. Since the potential is constant in different tunneling regions, it is possible to make an ansatz as super position of plane-waves
\begin{equation}
\widetilde\Psi(x)=u(k_x,k_y)\,e^{ik_x x}.
\end{equation}
Then the stationary equation for each region becomes
\begin{equation}
H(k_x,k_y)\,u(k_x,k_y)=E\,u(k_x,k_y).
\end{equation}
For the rotationally symmetric Rarita-Schwinger Hamiltonian we recall that we have chosen $v_2=0$. The equation then yields two propagating branches, which we label by their spin projection, $m=1/2,\qquad m=3/2$. Interpreting the problem as a generalized eigenvalue problem for $k_x$ at a fixed energy $E$, potential $V_0$ and transverse momentum $k_y$, we directly find the corresponding longitudinal wave numbers $k_x$. We denote the result in free regions ($V=0$) as $k_i$ and inside the barrier region ($V=V_0$) as $q_i$. The results are given below
\begin{equation}
k_{m}=\sqrt{\frac{E^2}{m^2v_1^2}-k_y^2},\quad q_{m}=\sqrt{\frac{(E-V_0)^2}{m^2v_1^2}-k_y^2},
\label{eq:k_out_clean}
\end{equation}
where we stress that index $m\in\{1/2,3/2\}$ corresponds to the respective spin projection sector.

It is convenient to parameterize the direction of propagation in each region by
\begin{equation}
\phi_m=\arctan(k_m,k_y),
\quad
\theta_m=\arctan(q_m,k_y),
\label{angles}
\end{equation}
where we used the version of $\arctan$ that has two entries to ensure correct treatment of different angle sectors. We note to avoid confusion about order of entries that $\arctan(a,b)=\mathrm{atan2}(b,a)$ in some languages like Python.
In this notation, the free-region eigenspinors may be chosen as $u_m^{\sigma}(\phi)$, where $\sigma=\pm$ denotes right-moving "$+$" and left-moving "$-$" solutions and $m$ the spin projection sector. Explicit expressions are given below
\begin{equation}
u_{1/2}^{\sigma}(\phi)=
\begin{pmatrix}
-s_\sigma\, e^{-3is_\sigma\phi}\\
-\frac{1}{\sqrt3}e^{-2is_\sigma\phi}\\
\frac{s_\sigma}{\sqrt3}e^{-is_\sigma\phi}\\
1
\end{pmatrix},
\quad
u_{3/2}^{\sigma}(\phi)=
\begin{pmatrix}
s_\sigma\, e^{-3is_\sigma\phi}\\
\sqrt3\,e^{-2is_\sigma\phi}\\
\sqrt3\,s_\sigma\,e^{-is_\sigma\phi}\\
1
\end{pmatrix},
\label{eq:spinors_clean}
\end{equation}
where we set $s_\pm=\pm1$.
With this notation, the scattering states in the three regions take the relatively compact form
\begin{equation}
\begin{aligned}
\widetilde\Psi_{\mathrm I}(x)
&=
\delta_{1/2}\,u_{1/2}^{+}(\phi_{1/2})\,e^{ik_{1/2}x}
+r_{1/2}\,u_{1/2}^{-}(\phi_{1/2})\,e^{-ik_{1/2}x}
\nonumber\\
&+
\delta_{3/2}\,u_{3/2}^{+}(\phi_{3/2})\,e^{ik_{3/2}x}
+r_{3/2}\,u_{3/2}^{-}(\phi_{3/2})\,e^{-ik_{3/2}x},
\\
\widetilde\Psi_{\mathrm {II}}(x)
&=
a_{1/2}^{+}\,u_{1/2}^{+}(\theta_{1/2})\,e^{iq_{1/2}x}
+a_{1/2}^{-}\,u_{1/2}^{-}(\theta_{1/2})\,e^{-iq_{1/2}x}
\nonumber\\
&+
a_{3/2}^{+}\,u_{3/2}^{+}(\theta_{3/2})\,e^{iq_{3/2}x}
+a_{3/2}^{-}\,u_{3/2}^{-}(\theta_{3/2})\,e^{-iq_{3/2}x},
\\
\widetilde\Psi_{\mathrm {III}}(x)
&=
t_{1/2}\,u_{1/2}^{+}(\phi_{1/2})\,e^{ik_{1/2}x}
+t_{3/2}\,u_{3/2}^{+}(\phi_{3/2})\,e^{ik_{3/2}x}.
\end{aligned}
\label{eq:psiIII_clean}
\end{equation}
Here, the $\delta_m$ denote incoming amplitudes, $r_m$ reflected amplitudes, $a_m^\pm$ the barrier-region amplitudes, and $t_m$ the transmitted amplitudes. The two incoming sectors belong to the same spin-$3/2$ multiplet. For coherent mixed incidence, we assume a fixed
relative phase between $\delta_{1/2}$ and $\delta_{3/2}$, as could
arise when a single incoming contact mode couples to both branches.
A contact-specific preparation model is beyond the present analysis.

Because the Hamiltonian is first order in spatial derivatives, the boundary conditions require only continuity of the four-component spinor:
\begin{equation}
\Psi_{\mathrm I}(0,y)=\Psi_{\mathrm {II}}(0,y),
\quad
\Psi_{\mathrm {II}}(L,y)=\Psi_{\mathrm {III}}(L,y),
\label{eq:bc_clean}
\end{equation}
which provide eight linear equations for the eight unknown coefficients $r_{1/2},\,r_{3/2},\,a_{1/2}^{\pm},\,a_{3/2}^{\pm},\,t_{1/2},\,t_{3/2}.$
In practice, one simply solves this mode-matching problem. If desired, the result can be written in a compact matrix form by eliminating the barrier amplitudes,
\begin{equation}
\begin{pmatrix}
\delta_{1/2}\\
r_{1/2}\\
\delta_{3/2}\\
r_{3/2}
\end{pmatrix}
=
\mathcal N
\begin{pmatrix}
t_{1/2}\\
0\\
t_{3/2}\\
0
\end{pmatrix},
\label{eq:N_clean}
\end{equation}
with a $4\times4$ matrix $\mathcal N$ determined by the two interfaces. This matrix is just a compact way of summarizing the same boundary matching and need not be written out explicitly in the main text.  A more detailed derivation of expressions has been relegated to appendix \ref{appendix}. We stress that using a treatment with matrix $\mathcal N$ has certain advantages for multi-barrier systems - matrices can be chained. While we will not be treating multi-barrier systems in the present work, we kept this language to help facilitate possible future work.  Solving Eq. \eqref{eq:N_clean} for the transmission amplitudes $t_m$ gives
\begin{align}t_{3/2}
&=
\frac{\mathcal N_{11}\delta_{3/2}-\mathcal N_{31}\delta_{1/2}}
{\mathcal N_{11}\mathcal N_{33}-\mathcal N_{13}\mathcal N_{31}},\quad t_{1/2}=
\frac{\mathcal N_{33}\delta_{1/2}-\mathcal N_{13}\delta_{3/2}}
{\mathcal N_{11}\mathcal N_{33}-\mathcal N_{13}\mathcal N_{31}}.
\label{eq:t12_clean}
\end{align}
The expressions make a basic physical fact of the system clear: even if only one sector is incident, the transmitted state generally contains both outgoing sectors, because the barrier mixes the two propagating branches.

Next, we want to obtain the transmission probability. To be careful in our treatment of this multi-component tunneling problem, we don't just compute $|t_m|^2$ like a single component tunneling case would suggest. Rather, we start our discussion directly from first principles with the longitudinal current operator 
\begin{equation}
\hat J_x = v_1 J_x.
\end{equation}
For any spinor $\Psi$, the corresponding longitudinal current density is
$J_x[\Psi]= \Psi^\dagger \hat J_x \Psi.$
Evaluating this current on-shell \footnote{Momenta and energies are restricted to what is allowed by the dispersion relation.} with the spinor basis $u_m^\pm$ yields
\begin{equation}
(u_m^{\pm}(\phi_m))^\dagger J_x u_m^{\pm}(\phi_m)
=
\pm \frac{16}{3}\,m^2\cos\phi_m.
\end{equation}
More importantly, we find that the current is \emph{diagonal on-shell} in our basis as seen below
\begin{equation}
u_{1/2}^{\sigma}(\phi_{1/2})^\dagger J_x u_{3/2}^{\tau}(\phi_{3/2})=0=
u_{m}^{+}(\phi_m)^\dagger J_x u_{m}^{-}(\phi_m),
\label{eq:Jx_orth_clean}
\end{equation}
where $\sigma,\tau=\pm$, provided $\phi_{1/2}$ and $\phi_{3/2}$ correspond to the same fixed pair $(E,k_y)$. Thus, the transmitted current has no interference term between the two branches, even though the transmitted wave function is, in general, a coherent superposition of them.

After a short computation, it then follows that the incident and transmitted currents are
\begin{align}
J_x^{\rm inc}
&=
v_1\left[
\frac{4}{3}|\delta_{1/2}|^2\cos\phi_{1/2}
+
12|\delta_{3/2}|^2\cos\phi_{3/2}
\right],\label{eq:JincJinc_clean}
\\
J_x^{\rm tra}
&=
v_1\left[
\frac{4}{3}|t_{1/2}|^2\cos\phi_{1/2}
+
12|t_{3/2}|^2\cos\phi_{3/2}
\right].
\label{eq:JincJtra_clean}
\end{align}
Therefore, the exact branch-resolved transmission probabilities are
\begin{equation}
T_m=
\frac{m^2\cos\phi_m\,|t_m|^2}
{\frac{1}{4}|\delta_{1/2}|^2\cos\phi_{1/2}
+
\frac{9}{4}|\delta_{3/2}|^2\cos\phi_{3/2}},
\label{eq:transmission_spin_resolved}
\end{equation}
and the total transmission is
\begin{equation}
T = \frac{J_x^{\rm tra}}{J_x^{\rm inc}} = T_{1/2}+T_{3/2}.
\label{eq:Ttotal_clean}
\end{equation}
This formula is exact. We also emphasize that the result is more complicated than just a $|t_m|^2$ as one would naively have expected. 

However, we may also directly note that expressions simplify for single-channel incidence. Here, one sets either $\delta_{3/2}=0$ or $\delta_{1/2}=0$, but the transmitted current still contains both $t_{1/2}$ and $t_{3/2}$, because scattering at the barrier can convert one branch into the other. Two  especially simple limits are worth mentioning. First, if only one branch is injected, then Eq.~\eqref{eq:transmission_spin_resolved} reduces to
\begin{equation}
    T_m=\frac{\left|t_m\right|^2}{\left|\delta_m\right|^2},
\end{equation}
where $i\in\{1/2,3/2\}$. Second, at normal incidence $k_y=0$, one has
\begin{equation}
\phi_{1/2}=\phi_{3/2}=\theta_{1/2}=\theta_{3/2}=0,
\end{equation}
so all angle-dependent spinor phases disappear in Eqs.~\eqref{eq:JincJinc_clean}--\eqref{eq:JincJtra_clean} and the transmission formulas simplify to
\begin{equation}
\begin{aligned}
    &J_x^{\rm inc}=v_1\left(\frac43|\delta_{1/2}|^2+12|\delta_{3/2}|^2\right)\\
&J_x^{\rm tra}=v_1\left(\frac43|t_{1/2}|^2+12|t_{3/2}|^2\right).
\end{aligned}
\end{equation}
One may show that, for normal incidence, $|\delta_{i}|=|t_{i}|$ and hence
\begin{equation}
    T= \frac{J_x^{\rm tra}}{J_x^{\rm inc}} =1
\end{equation}
and therefore obtain Klein tunneling.

\subsection{Results and discussion}
We next turn our discussion to a comparative investigation of transmission properties of the barrier in two situations. First, we consider incidence from a single spin-projection sector, which allows us to compare the transport properties of the $m=1/2$ and $m=3/2$ branches separately - similar to existing literature \cite{MANDAL2020126666,Mandal2023}. Second, we consider an incoming wave that is a coherent mixture of both sectors. 
In what follows, we examine both the over-barrier regime ($E/V_0=1.5$) and the tunneling regime ($E/V_0=0.8$). Here, energy values were chosen as prototypical examples of each regime. In all plots, we use dimensionless conserved transverse momentum
\begin{equation}
\kappa_y \equiv \frac{v_1 k_y}{V_0}.
\label{eq:kappay}
\end{equation}
instead of an angle. This choice for our purposes is preferable to an angular axis because for a fixed $k_y$ and energy $E$, the two propagating branches correspond to different physical incidence angles, $\phi_{1/2}$ and $\phi_{3/2}$. This would complicate a comparison.

\begin{figure}[ht!]
\centering
    \includegraphics[width=0.48\columnwidth]{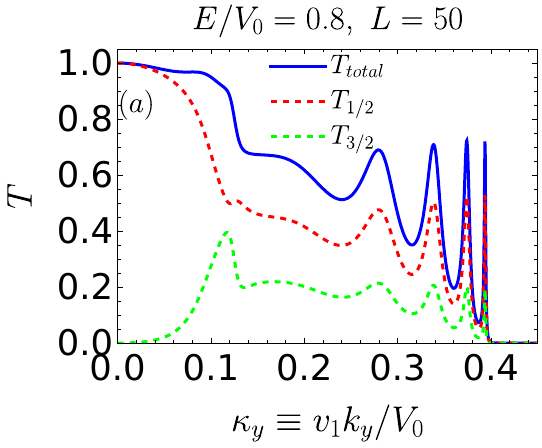}
    \includegraphics[width=0.48\columnwidth]{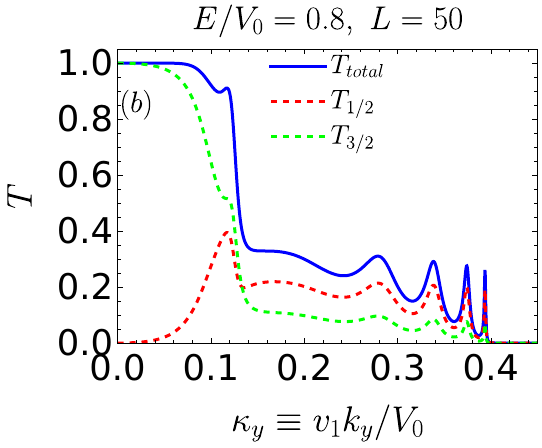}
    \includegraphics[width=0.48\columnwidth]{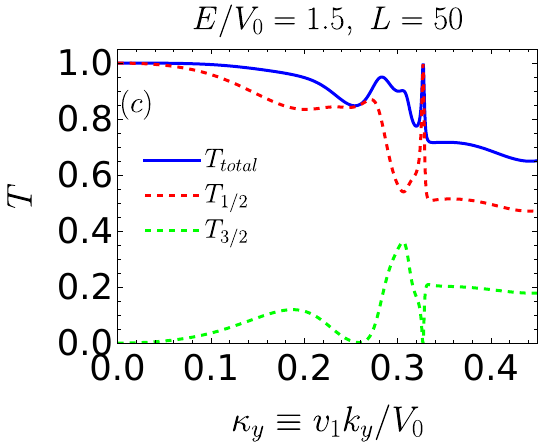}
    \includegraphics[width=0.48\columnwidth]{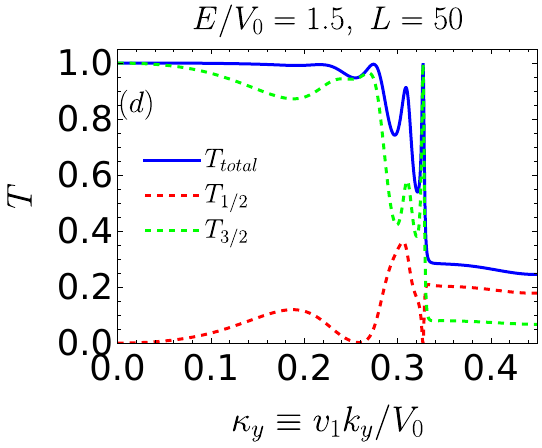}
\caption{Transmission probabilities of spin 1/2, 3/2 and the total for independent incoming channels.  The upper (a,b) panels show the tunneling regime $E/V_0=0.8$, while the lower (c,d) panels show the over-the-barrier regime $E/V_0=1.5$. The left panels (a,c) correspond to incoming spin projection $m=1/2$ with $\delta_{1/2}=1$ and $\delta_{3/2}=0$, while the right (b,d) panels correspond to incoming spin projection $m=3/2$ with $\delta_{3/2}=1$ and $\delta_{1/2}=0$. The horizontal axis is the dimensionless conserved transverse momentum $\kappa_y=v_1k_y/V_0$. We included only positive $\kappa_y$ because negative $\kappa_y$ corresponds to a mirror image.} 
\label{fig:single_modes}
\end{figure}

We begin with independent incidence in a single sector. Here,
Fig.~\ref{fig:single_modes}(a,b) corresponds to a genuine tunneling regime and Fig.~\ref{fig:single_modes}(c,d) shows the over-barrier case. 
We find that, for each incoming channel, the transmission is an even function of $\kappa_y$, as expected for a mirror-symmetric barrier with rotationally symmetric bulk dispersion. Moreover, both channels (left for spin $1/2$ and right for spin $3/2$) exhibit Klein tunneling $T(0)=1$ in agreement with the perfect-transmission result at normal incidence derived in the previous section \cite{ZHU2026170301,MANDAL2020126666}. We see that the spin projection is conserved for normal incidence. 
Away from normal incidence, however, the two spin projection sectors mix and show behavior dependent on barrier height. In the over-barrier regime, Fig.~\ref{fig:single_modes}(c,d), our most striking observation is near the edge of the plot where the $m=3/2$ channel develops slightly stronger oscillations and deeper minima than the $m=1/2$ case. In the tunneling regime, Fig.~\ref{fig:single_modes}(a,b), transmission away from normal incidence is suppressed much more strongly. The $m=1/2$ sector at large $|\kappa_y|$ corresponds to the most opaque barrier setting of the present figures. The $m=3/2$ sector remains comparatively more transmissive for the same parameter range. A possible explanation of this effect is related to the different propagation speeds $v_1/2$ and $3v_1/2$ for different spin projection spinors. Here, the idea is that the $m=3/2$ sector spends less time inside the barrier and hence decays less.  

\begin{figure}[ht]
\centering
     \includegraphics[width=0.48\columnwidth]{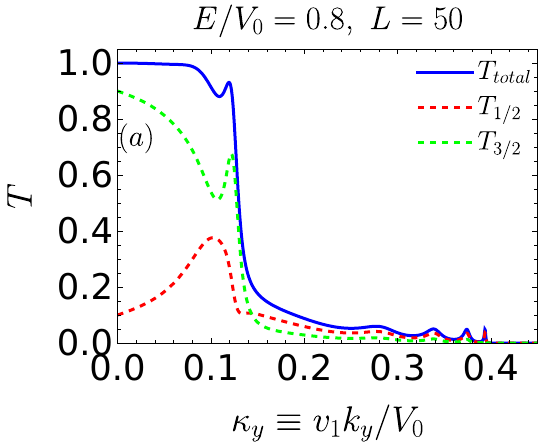}
    \includegraphics[width=0.48\columnwidth]{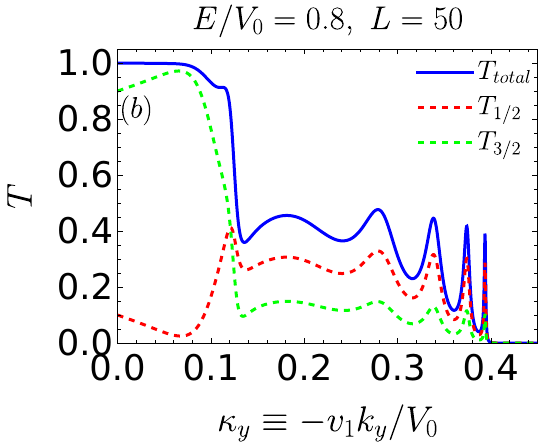}
    \includegraphics[width=0.48\columnwidth]{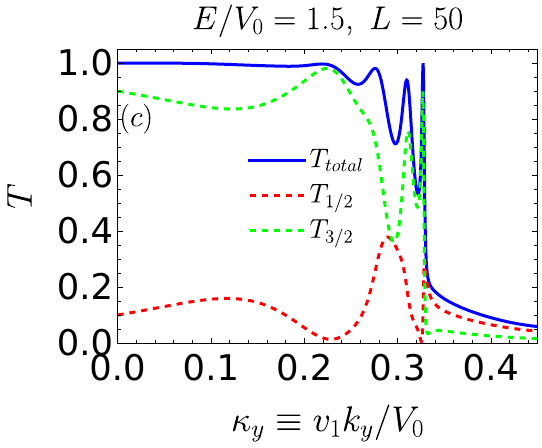}
    \includegraphics[width=0.48\columnwidth]{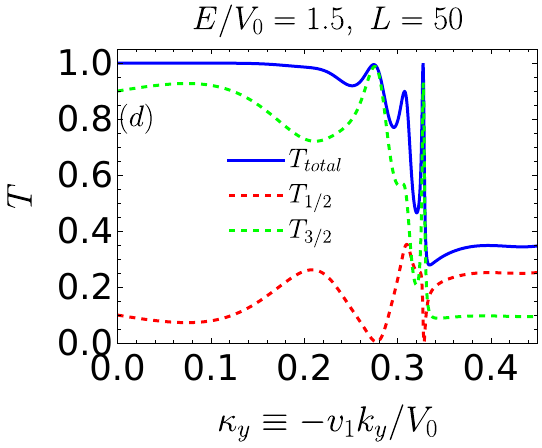}
\caption{Transmission for a coherent equal-amplitude superposition with $\delta_{1/2}=\delta_{3/2}=1$. The top row  corresponds to the tunneling regime $E/V_0=0.8$, and the bottom row corresponds to the over-barrier regime $E/V_0=1.5$. The left and right columns in the figure correspond to $T(\kappa_y)$ and $T(-\kappa_y)$, respectively.}
\label{fig:coherent_modes}
\end{figure}

We next turn to the case of an incident coherent mixture of both sectors in Fig.~\ref{fig:coherent_modes} below. To make the calculation definite, we choose equal incoming amplitudes with zero relative phase; specifically we choose $\delta_{1/2}=\delta_{3/2}=1$.
In Fig.~\ref{fig:coherent_modes}, we demonstrate broken mirror symmetry by plotting  $T(\kappa_y)$ (the left column) and $T(-\kappa_y)$ (the right column). In both cases, the two curves coincide at normal incidence, but then separate away from $\kappa_y=0$, which indicates that the tunneling transmission is no longer an even function in $\kappa_y$. We observe that the effect is increasingly pronounced at larger incident angle (or equivalently, its proxy parameter $\kappa_y$). It therefore appears only once the incoming state probes the $k_y$-dependent spinor structure of the problem. We also note that the effect is much more pronounced in the tunneling regime of Fig. \ref{fig:coherent_modes}(a,b) than in the over-the-barrier regime in Fig. \ref{fig:coherent_modes}(c,d). This likely reflects the reduced influence of the barrier in the over-barrier regime. We also note that the near-unity transmission peaks at finite $\kappa_y$ in
Figs.~\ref{fig:single_modes} and \ref{fig:coherent_modes} are
barrier-dependent Fabry--P\'erot-type transmission resonances,
distinct from the normal-incidence Klein tunneling discussed above.

\begin{figure}[ht]
\centering
     \includegraphics[width=0.48\columnwidth]{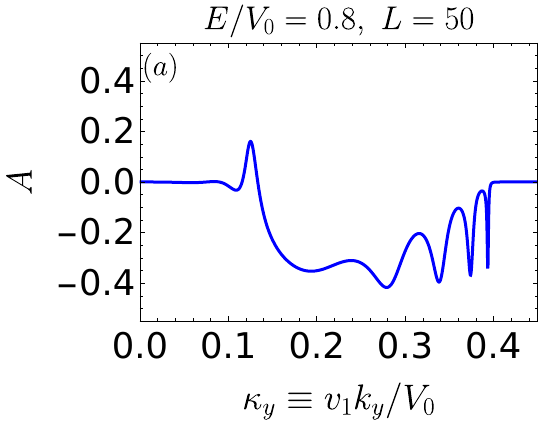}
    \includegraphics[width=0.48\columnwidth]{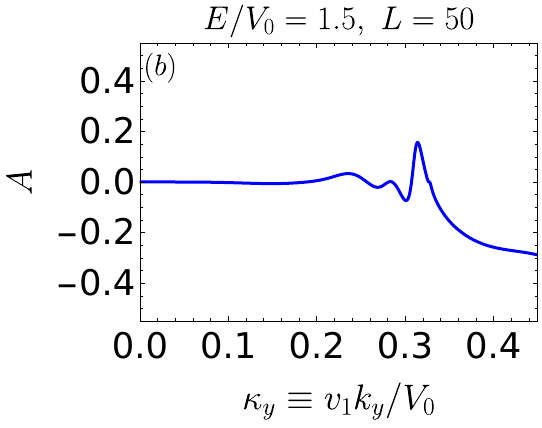}
  
\caption{The asymmetry function as represented in Eq.~\eqref{equ:Asy} for a coherent equal-amplitude superposition with $\delta_{1/2}=\delta_{3/2}=1$. The left column corresponds to the tunneling regime $E/V_0=0.8$, and the right column to the over-barrier regime $E/V_0=1.5$ }
\label{fig:coherent_modes_Asy}
\end{figure}

To better quantify the directional dependence in Fig.~\ref{fig:coherent_modes}, we also plotted the asymmetry function
\begin{equation}
A(\kappa_y)= T(\kappa_y)-T(-\kappa_y)
\label{equ:Asy}
\end{equation}
in Fig.~\ref{fig:coherent_modes_Asy}, which by construction becomes zero for a perfectly symmetric case. These plots provide a clear demonstration of an emergent asymmetry. We observe in both transmission regimes ( $E/V_0=1.5$ and $E/V_0=0.8$) that $A(\kappa_y)$ develops increasingly pronounced peaks away from normal incidence. We therefore conclude that the directional response is not an artifact of the regimes $E>V_0$ or $E<V_0$.

In summary, the three figures \ref{fig:single_modes}, \ref{fig:coherent_modes} and \ref{fig:coherent_modes_Asy} highlight two qualitatively distinct effects. For independent incidence, the two branches display quantitatively different transmission profiles, but remain exactly symmetric under $\kappa_y\to-\kappa_y$. In contrast, a coherent superposition of incident channels leads to direction-dependent transmission, with asymmetry persisting above and below the barrier. At first sight, the directional response appears surprising due to the barrier symmetry and rotationally symmetric dispersion. In the next section, we will see that it originates from coherent coupling between the two spin-projection sectors together with the $k_y$-dependent spinor structure of the scattering states, rather than from any explicit geometric asymmetry of the barrier or dispersion relation.

\section{Explanation of asymmetry and relation to  Imbert-Fedorov effect}
\label{Explanation}

In this section, we explain why the transmission is symmetric for a single incident channel, but can become asymmetric under $\kappa_y\to-\kappa_y$ for a coherent superposition of the two propagating sectors. The central point is that the asymmetry does not originate from an anisotropy of the energy spectrum. Indeed, in the rotationally symmetric case considered here, the spectrum depends only on $k_x^2+k_y^2$, so changing $k_y\to -k_y$ leaves the energies unchanged. The nontrivial part lies instead in the \emph{spinor structure} of the scattering states.

We first begin with an intuitive but incomplete picture. As seen in Eq.~\eqref{eq:spinors_clean}, the propagating eigenspinors contain phases
\begin{equation}
e^{\pm i\phi_m},\quad e^{\pm 2i\phi_m},\quad e^{\pm 3i\phi_m},
\end{equation}
and the barrier-region spinors contain the analogous phases built from $\theta_m$. Clearly, if we replace $k_y\to -k_y$, the longitudinal wave numbers $k_m$ and $q_m$ remain unchanged. However, as seen from Eq.~\eqref{angles}, the angles change sign with
\begin{equation}
\phi_m(-k_y)=-\phi_m(k_y),\quad
\theta_m(-k_y)=-\theta_m(k_y).
\end{equation}
Therefore, the wave functions do not transform trivially under reflection: the spectrum is even in $k_y$, but the spinors pick up nontrivial phases.
These phases enter directly into the interface-matching problem and, therefore, into the transmission amplitudes. In addition, the two propagating spin projection sectors $m=1/2,3/2$ accumulate different dynamical phases across the barrier because they propagate with different velocities $v_m=mv_1$ (slope of the Dirac cone). More formally, this accumulation of phase difference can also be seen from the fact that in general $q_{1/2}\neq q_{3/2}$. This is because the dynamic phase difference is $\Delta \Phi^{\mathrm{dyn}}=\left(q_{3 / 2}-q_{1 / 2}\right) L$ for a barrier of thickness $L$. The observed directional effect then is a consequence of this phase structure together with coherent mixing between the two sectors.

Having provided an intuitive picture, we now turn to a more formal discussion. First, we observe that because the scattering problem is linear, each transmitted amplitude is a linear combination of the two incoming amplitudes
\begin{equation}
t_m(k_y)=A_m(k_y)\,\delta_{1/2}+B_m(k_y)\,\delta_{3/2}.
\label{eq:tm_linear}
\end{equation}
Here $A_m$ and $B_m$ are determined by the transfer matrix $\mathcal{N}$ (see Eq. \eqref{eq:t12_clean}) and inherit their $k_y$-dependence from the spinor phases and the barrier phases discussed above. In other words, the coefficients $A_m$ and $B_m$ are explicit, although algebraically complicated, functions of
\begin{equation}
e^{\pm i\phi_{1/2}},\ e^{\pm i\phi_{3/2}},\
e^{\pm i\theta_{1/2}},\ e^{\pm i\theta_{3/2}},\
e^{\pm iq_{1/2}L},\ e^{\pm iq_{3/2}L}.
\end{equation}
The total transmission, we recall is $T=T_{1/2}+T_{3/2}$
with $T_m$ given in Eq.~\eqref{eq:transmission_spin_resolved}. Importantly, the transmitted current is diagonal on shell in the basis of propagating branches, as stated in Eq.~\eqref{eq:Jx_orth_clean}. Thus the asymmetry does \emph{not} come from direct interference between the transmitted $m=1/2$ and $m=3/2$ branches in the \emph{outgoing flux}. Instead, it comes from interference within each term $|t_m|^2$ because each outgoing branch can be fed by two distinct incident sectors.

Before discussing the asymmetric case, let us first suppose that only one sector is incident, for example
\begin{equation}
\delta_{3/2}=0.
\end{equation}
Then Eq.~\eqref{eq:tm_linear} reduces to
\begin{equation}
t_m(k_y)=A_m(k_y)\,\delta_{1/2},
\end{equation}
and therefore
\begin{equation}
|t_m(k_y)|^2=|A_m(k_y)|^2\,|\delta_{1/2}|^2.
\end{equation}
There is no mixed term between the two incident sectors because only one sector is present. After taking the modulus squared, all phase factors ($e^{\pm n\phi_m}$ etc.) in $A_m(k_y)$ appear paired with their complex conjugates. As a result, the dependence on $k_y$ enters only through even combinations such as cosine-type terms, and the transmission remains symmetric
\begin{equation}
T(k_y)=T(-k_y).
\end{equation}
This explains why the single-channel curves in Fig.~\ref{fig:single_modes} are even functions of $k_y$.

Next, we turn to the important case of coherent two-channel incidence.
Here, the situation changes qualitatively when both sectors are injected coherently. In that case
\begin{equation}
t_m(k_y)=A_m(k_y)\,\delta_{1/2}+B_m(k_y)\,\delta_{3/2},
\end{equation}
and hence
\begin{align}
|t_m(k_y)|^2
&=
|A_m(k_y)|^2\,|\delta_{1/2}|^2
+
|B_m(k_y)|^2\,|\delta_{3/2}|^2
\nonumber\\
&\quad
+
2\,\mathrm{Re}\!\left[
A_m(k_y)B_m^*(k_y)\,\delta_{1/2}\delta_{3/2}^*
\right].
\label{eq:interference_term}
\end{align}
The last term is the \emph{crucial new contribution}. Since $A_m$ and $B_m$ carry different $k_y$-dependent phases, this interference term need not be even under $k_y\to -k_y$. We emphasize that equation~\eqref{eq:interference_term} also shows that merely populating both incident sectors is insufficient to produce the asymmetry, a coherent superposition is needed. After removing an irrelevant overall phase, an incoherent mixture may be represented by taking $\delta_{1/2}=|\delta_{1/2}|$ and $\delta_{3/2}=|\delta_{3/2}|e^{i\chi}$ and averaging over a random relative phase $\chi$. Since
\begin{equation}
\left\langle \delta_{1/2}\delta_{3/2}^{*}\right\rangle_{\chi}
=
|\delta_{1/2}||\delta_{3/2}|
\left\langle e^{-i\chi}\right\rangle_{\chi}
=0,
\end{equation}
the interference term in Eq.~\eqref{eq:interference_term} vanishes. The remaining transmission is an incident-current-weighted combination of the two single-channel transmissions. Because each single-channel transmission is even under $k_y\to-k_y$, the phase-averaged transmission of the incoherent mixture satisfies
\begin{equation}
\overline{T}(k_y)
\equiv
\left\langle T(k_y;\chi)\right\rangle_{\chi}
=
\left\langle T(-k_y;\chi)\right\rangle_{\chi}
\equiv
\overline{T}(-k_y).
\end{equation}
Thus, we stress that the directional asymmetry requires a well-defined relative phase between the two incident sectors and therefore occurs only for a coherent superposition.

Now that an understanding of the effect based on tunneling coefficients has been established, we want to understand the effect better fundamentally from a symmetry perspective. Here, we turn our attention to a discussion of the mirror symmetry of the Hamiltonian in the $y$-direction. Specifically, the asymmetry we observe does not imply that mirror symmetry is broken. Indeed, reflection in the $y$-direction is represented by a unitary operator
\begin{equation}
U_y=e^{-i\pi J_x}.
\end{equation}
Specifically, one may observe that
\begin{equation}
U_y J_x U_y^{-1}=J_x,
\quad
U_y J_y U_y^{-1}=-J_y.
\end{equation}
Hence, the Hamiltonian satisfies
\begin{equation}
U_y\,H(k_x,k_y)\,U_y^{-1}=H(k_x,-k_y)
\label{eq:mirror_H_short}.
\end{equation}
This confirms that we have found the appropriate mirror-symmetry operator.

The important point to observe here is that mirror symmetry acts non-trivially on the propagating spinors, as shown below
\begin{equation}
U_y\,u_m^{+}(\phi_m)
=
i(2m-2)\,e^{-3i\phi_m}\,u_m^{+}(-\phi_m).
\label{eq:mirror_spinors_short}
\end{equation}
Thus, under $k_y\to -k_y$, the two incoming sectors acquire different phase factors. For single-channel incidence, this produces only an overall phase and therefore does not affect the transmission probability, so $T(k_y)=T(-k_y)$. For coherent mixed incidence, however, the relative phase between the two sectors changes under reflection. Since each transmitted branch is fed by both incoming sectors, this relative phase enters the interference term in Eq.~\eqref{eq:interference_term}. Consequently, the transmission for a fixed coherent input can satisfy
\begin{equation}
T(k_y)\neq T(-k_y),
\end{equation}
without any contradiction with the mirror symmetry of the Hamiltonian or of the barrier.

Next, we want to relate our result to a well-known optical effect. The above mechanism suggests a natural analogy to the so-called Imbert-Fedorov effect in optics. There, different polarization components acquire different phases upon reflection or refraction at an interface, and for a finite beam, this phase difference manifests itself as a transverse shift. In the present problem, the two propagating sectors $m=1/2$ and $m=3/2$ play a role similar to that of polarization as internal degrees of freedom that are transformed differently by the barrier. For single-channel incidence, this has no observable consequence beyond an overall phase. For coherent mixed incidence, however, the relative phase between the two sectors enters the interference term in Eq.~\eqref{eq:interference_term}, making the transmission direction dependent. The analogy with the Imbert--Fedorov effect is therefore limited to the common role of internal-state-dependent interface phases. Our plane-wave calculation does not establish a real-space transverse displacement of a wavepacket; rather, it reveals an analogous phase-driven asymmetry in the transmission probability. This analogy should not be interpreted as a statement about Berry-curvature-induced anomalous velocity \cite{PhysRevLett.93.083901,PhysRevLett.115.156602,PhysRevB.96.115448}.

The requirement of in incident wave that is a coherent of spin projection sectors mixture naturally raises the question of its experimental preparation. Coherent optical manipulation of electronic spin states and electrical injection of spin-polarized carriers have already been demonstrated in semiconductor systems \cite{Gupta2001SpinCoherence,Fiederling1999SpinInjection,Ohno1999SpinInjection}. Moreover, optical transitions involving spin-$3/2$ multifold bands have been studied theoretically \cite{SanchezMartinez2019Optical}. These results suggest that a coherent mixture could, in principle, be prepared by phase-coherent optical excitation or by a suitably engineered spin-selective contact or interface whose injected state has coherent overlap with both propagating sectors. We emphasize, however, that the preparation of such a controlled superposition in a Rarita--Schwinger semimetal remains to be demonstrated experimentally. Observation of the directional asymmetry requires the relative phase to remain stable during barrier traversal. Dephasing, phase-randomized injection, or ensemble averaging suppresses the interference contribution and restores mirror-symmetric transmission.

The main lesson from our work is that directional tunneling asymmetries in multiband systems do not necessarily arise from explicit band anisotropy. In the present rotationally symmetric model, the effect already appears because the barrier acts nontrivially on the internal structure of the wave function and because coherent superpositions of the two propagating sectors allow the resulting phase difference to become observable in the transmission. In more realistic materials, additional crystal anisotropies may, of course, also contribute. Our results, however, show that asymmetric tunneling can already arise in an idealized isotropic model and should not be automatically attributed solely to anisotropic dispersion. It is therefore safer to consider both effects.
\section{Macroscopic observables: momentum-integrated responses}

We next connect the mode-resolved transmission probabilities to
macroscopic transport observables. Within linear-response Landauer
theory, the longitudinal conductance per unit transverse length is
\cite{PhysRevB.31.6207,PhysRevB.40.1456}
\begin{equation}
\mathcal{G}_{x}
\equiv \frac{G_x}{W}
=
\frac{e^2}{2\pi h}
\int dE
\left(-\frac{\partial f}{\partial E}\right)
\int_{\mathcal D(E)} dk_y\,
T(E,k_y),
\label{eq:longitudinal_conductance}
\end{equation}
where $f(E)$ is the Fermi--Dirac distribution and $\mathcal D(E)$
denotes the transverse-momentum interval over which the relevant
incident modes propagate. This interval is symmetric under
$k_y\to-k_y$.

To examine the effect of the transmission asymmetry, we separate the parts of the transmission with different parity under
$k_y\to-k_y$ and write
\begin{equation}
T(E,k_y)
=
T_{\mathrm{s}}(E,k_y)
+
T_{\mathrm{a}}(E,k_y),
\end{equation}
where the symmetric $T_s$ and antisymmetric $T_a$ contributions are defined,
respectively, by
\begin{align}
T_{\mathrm{s}}(E,k_y)
&=
\frac{T(E,k_y)+T(E,-k_y)}{2}
\\
T_{\mathrm{a}}(E,k_y)
&=
\frac{T(E,k_y)-T(E,-k_y)}{2}.
\end{align}

At each fixed energy, the antisymmetric contribution $T_a$ is odd in $k_y$
and therefore integrates to zero over the symmetric interval
$\mathcal D(E)$. Consequently, only the symmetric contribution
$T_{\mathrm{s}}$ enters the longitudinal conductance and the asymmetric part of the transmission does
not contribute to the longitudinal conductance.

The situation is different for the transverse conductance carried by
the transmitted modes. Following the Landauer formulation for
transverse transport \cite{PhysRevB.110.205406}, its branch-diagonal
contribution per unit transverse length is
\begin{equation}
\begin{aligned}
\overline{\mathcal G}_{yx}^{\mathrm{tr}}
&\equiv
\frac{\overline{G}_{yx}^{\mathrm{tr}}}{W}=
\frac{e^2}{2\pi h}
\int dE
\left(-\frac{\partial f}{\partial E}\right)
\int_{\mathcal D(E)}dk_y\\
&\times\sum_{m=1/2,3/2}
\frac{v_{y,m}(E,k_y)}
     {v_{x,m}(E,k_y)}
T_m(E,k_y).
\label{eq:transverse_conductance}
\end{aligned}
\end{equation}
Here, $v_{i,m}=\partial E_m/\partial k_i$ is the group-velocity
component of branch $m$, where we continue to use $\hbar=1$. The overbar denotes spatial averaging over the observation coordinate
$x$, which removes interbranch interference contributions to the
transverse current that oscillate rapidly in space. Such contributions
are absent from the longitudinal current because the corresponding
off-diagonal matrix elements vanish on shell, as shown in
Eq.~\eqref{eq:Jx_orth_clean}. The overbar does not represent an
ensemble average over the relative phase of the two branches.

\begin{figure}
    \centering
    \includegraphics[width=1.\linewidth]{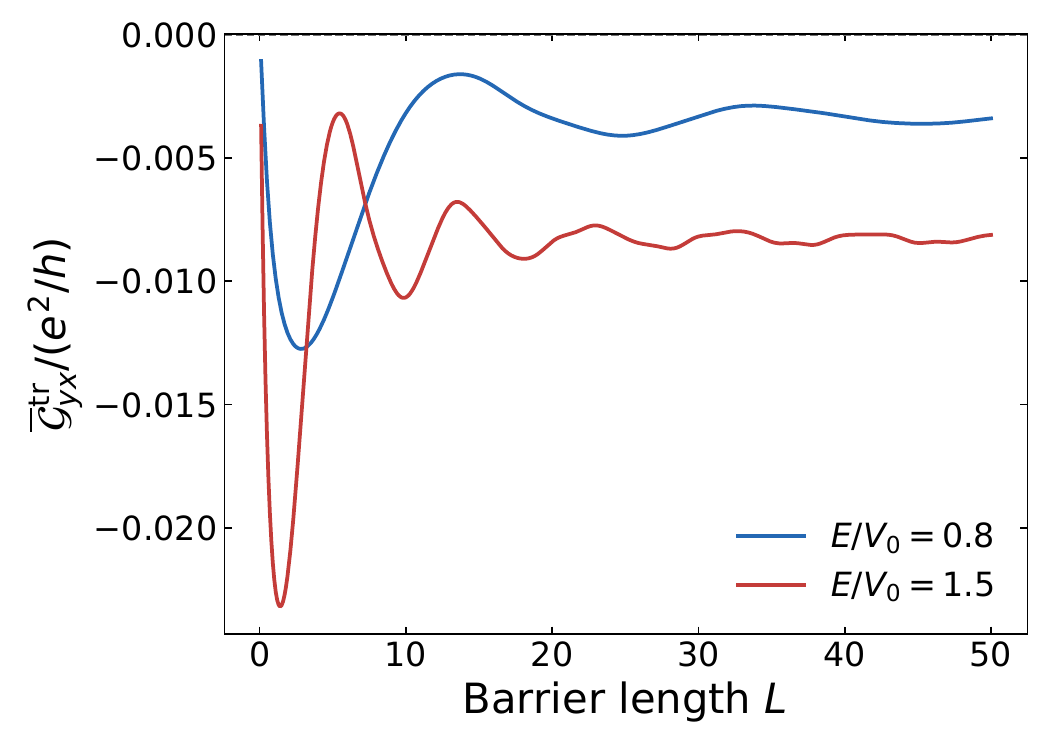}
    \caption{Plot of the transverse conductance for parameters $v_1=1$, $V_0=1$ and a coherent superposition of incoming waves $\delta_{1/2}=\delta_{3/2}=1/\sqrt{2}$. The blue curve corresponds to the tunneling regime with $E/V_0=0.8$ while the red curve corresponds to the over-the-barrier regime with $E/V_0=1.5$.}
    \label{fig:transverse_conductance}
\end{figure}

For the right-moving states,
\begin{equation}
v_{x,m}(-k_y)=v_{x,m}(k_y);
\quad
v_{y,m}(-k_y)=-v_{y,m}(k_y),
\end{equation}
so that $v_{y,m}/v_{x,m}$ is antisymmetric under $k_y\to-k_y$.
Consequently the symmetric part of $T_m(E,k_y)$ does not contribute to
Eq.~\eqref{eq:transverse_conductance}, whereas its antisymmetric part can produce a finite transverse transmitted-current response as we see in Fig. \ref{fig:transverse_conductance}.
The finite values of
$\overline{\mathcal G}_{yx}^{\mathrm{tr}}$ demonstrate that the
mode-resolved transmission asymmetry survives the momentum integration
and appears in the macroscopic transverse conductance. The oscillations with barrier length arise from coherent propagation and multiple reflection between the two barrier interfaces and may be viewed as multimode Fabry--Pérot-type interference. The observed resonances also identify barrier
lengths for which the magnitude of the transverse conductance is
enhanced. Within the present model, these lengths may help guide the choice of barrier geometry for experimental detection. They depend, however, on the incident energy, the prepared coherent
state, and the barrier parameters and are therefore not universal.
\section{higher Spin Bias}
\label{sec:higherspin_bias}

We now take a step back and investigate Figs.~\ref{fig:single_modes} and ~\ref{fig:coherent_modes} further. We recall that the plots show the wave transmission for two distinct energy regimes. These are the tunneling regime, $\frac{E}{V_0} < 1 $, and the scattering regime, $\frac{E}{V_0} > 1 $. The transmission is distinct as expected. However, in the over-the-barrier regime, the transmission seems to favor spin 3/2 (often, this component is larger). To further examine and clarify this effect, we define the transmission disparity
\begin{equation} \delta T = \frac{T_{3/2} - T_{1/2}}{T}, \label{equ:Tdis} \end{equation}
where $\delta T > 0$ corresponds to a case where the transmission of spin 3/2 is dominant and the opposite for $\delta T < 0$.

\begin{figure}[ht]
\centering
\includegraphics[width=0.95\columnwidth]{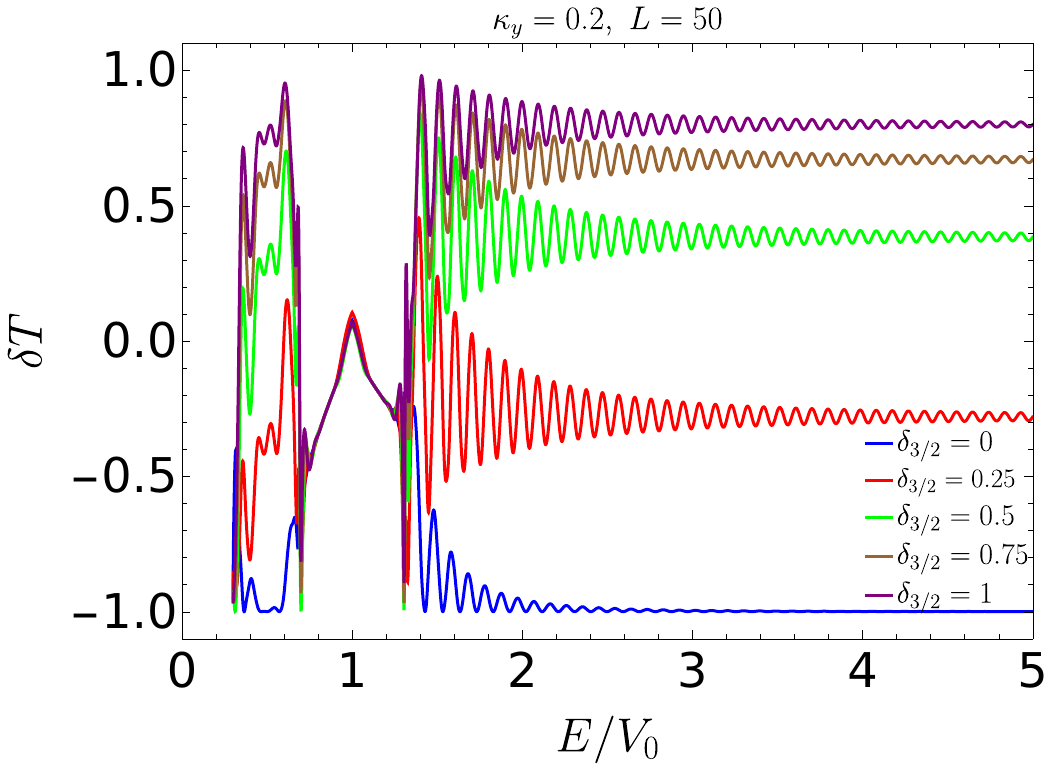}

\caption{Transmission disparity (Eq. \eqref{equ:Tdis}), for multiple initial coherent wave superpositions. All superpositions have fixed $\delta_{1/2} = 1$ and different values for $\delta_{3/2}$ represented by colored lines. }
\label{fig:Energy}
\end{figure}
First, in Fig.\ref{fig:Energy}, we investigate the transmission disparity as a function of energy $E/V_0$. This provides a foundation for the observations that follow.
We observe directly that the plots include curves that start at a finite non-zero energy. The reason for this is that energies below a certain threshold do not support the incoming transverse momentum $\kappa_y=0.2$ that was used on the plot (see Eqs. \eqref{eq:k_out_clean} and \eqref{eq:kappay}) and therefore starting the plots at lower energies was not possible. After establishing this, we next observe that for certain values of $E$ slightly below and above $V_0$, all curves coincide (see Fig.~\ref{fig:Energy}). This observation means that the transmission disparity becomes nearly universal across the tested coherent superpositions. This effect is not an artifact of a particular choice of parameters and appears more pronounced at larger values of $k_y$. More precisely, it corresponds to a regime in which barrier and spin-projection properties do not matter, and dynamics are decay-dominated. To see more clearly why this happens, we may observe that near $E=V_0$ or for large $k_y$, we find (see Eq. \eqref{eq:k_out_clean}) that $q_m\approx ik_y$, which is independent of the spin projection and leads to the decay. The independence of spin projection $m$ means that all coherent incoming mixtures behave similarly. In this regime, a bias towards higher-spin projections is not yet visible.
Next, we turn to high-energy behavior, which hints at a bias toward higher-spin transmission. Multiple curves, which correspond to smaller admixtures of spin 3/2 (i.e., $\delta_{3/2}<\delta_{1/2}$), have $\delta T>0$ in the large energy limit. This indicates more transmission into the $3/2$ channel, despite the smaller incoming contribution. The effect is most noticeable for the purple curve in Fig. ~\ref{fig:Energy}, where the outgoing wave is strongly dominated by spin $3/2$ with $\delta T\approx 0.8$. In contrast, unbiased transmission would have $\delta T = 0$.

\begin{figure}[ht]
\centering
\includegraphics[width=0.95\columnwidth]{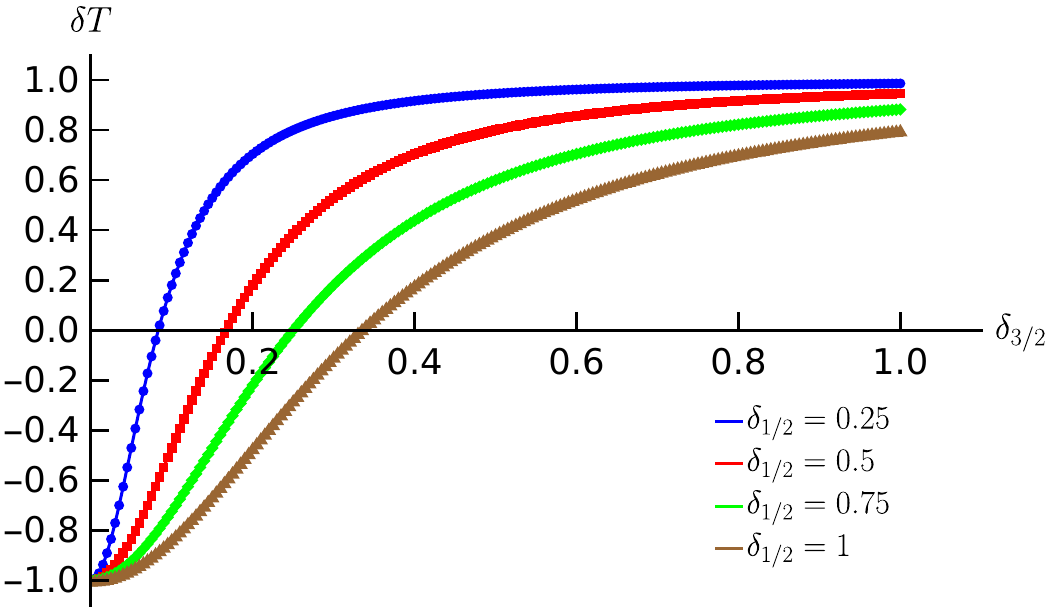}
\caption{The transmission disparity at high energy, $\delta T( E \sim 10^{4}  V_{0})$, versus $\delta_{3/2}$ for different values of $\delta_{1/2}$.}
\label{fig:AvHE}
\end{figure}
To further elucidate this bias, we now focus exclusively on the high-energy asymptotic behavior.
Fig. \ref{fig:AvHE} shows the asymptotic behavior of transmission disparity $\delta T$ as a function of $\delta_{3/2}$ for various fixed values of $\delta_{1/2}$.
As shown in Fig.~\ref{fig:AvHE}, the results provide a clearer indication that, at high-energy incidence, the system favors higher spin, as the curves are positive over most of the parameter range.

\begin{figure}[ht]
\centering
     \includegraphics[width=0.95\columnwidth]{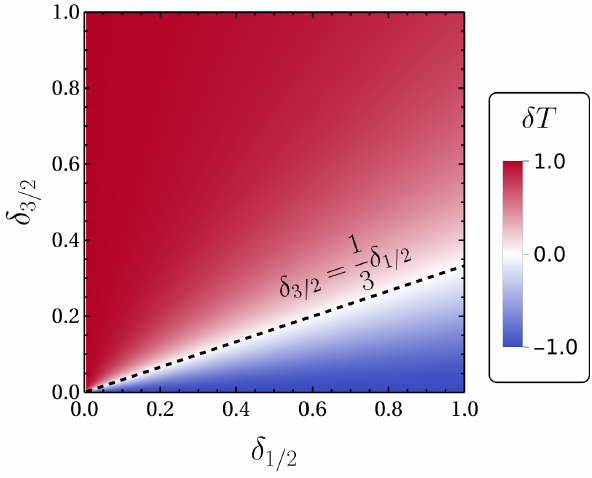}
\caption{A density plot of the transmission disparity at high energy, $\delta T( E \sim 10^{4}  V_{0})$ for all mixtures. The dashed line represents,  $\delta_{3/2} = \frac{1}{3}\delta_{1/2}$ with equal transmission amplitude,  $ \delta T $ = 0.}
\label{fig:all_Mixture}
\end{figure}
To gain even better insight, we next plotted $\delta T$ as a function of both mixture parameters $\delta_m$ (see Fig.~\ref{fig:all_Mixture}).
The figure shows that, for most mixtures, $\delta T>0$. Interestingly, we observe a linear relation between $\delta_{3/2}$ and $\delta_{1/2}$ for equal transmission amplitudes $\delta T=0$. Precisely we have that $\frac{\delta_{3/2}}{\delta_{1/2}} = \frac{1}{3}$. We therefore see that $1/6$ of the area is dominated by spin $1/2$, and the remaining $5/6$ by spin $3/2$, clearly showing a bias towards the $3/2$ projection in the transmission amplitude.

We can now analytically prove that $\frac{\delta_{3/2}}{\delta_{1/2}} = \frac{1}{3}$ corresponds to unbiased transmission in the high-energy limit, thereby strengthening our discussion. Indeed, in the large energy limit $t_m\approx \delta_m$. Then it is clear from Eq.~\eqref{eq:transmission_spin_resolved} and the definition of $\delta T$ that
\begin{equation}
\delta T(E\to\infty)=\frac{9\left|\delta_{3 / 2}\right|^2 \cos \phi_{3 / 2}-\left|\delta_{1 / 2}\right|^2 \cos \phi_{1 / 2}}{9\left|\delta_{3 / 2}\right|^2 \cos \phi_{3 / 2}+\left|\delta_{1 / 2}\right|^2 \cos \phi_{1 / 2}}.
\end{equation}
Moreover, we note that
\begin{equation}
\cos \phi_m=\sqrt{1-\frac{m^2 v_1^2 k_y^2}{E^2}} \rightarrow 1
\end{equation}
at high energies and therefore
\begin{equation}
\delta T \rightarrow \frac{9\left|\delta_{3 / 2}\right|^2-\left|\delta_{1 / 2}\right|^2}{9\left|\delta_{3 / 2}\right|^2+\left|\delta_{1 / 2}\right|^2}.
\end{equation}
It is now clear that coherent incident mixtures with $\frac{\delta_{3/2}}{\delta_{1/2}} = \frac{e^{i\nu}}{3}$ for any phase angle $\nu$ have unbiased transmission ($\delta T\to0$) in the high energy limit - extending our observation in the complex plane. Interestingly, the same approximations hold for a small incident angle (or equivalently, a small $\kappa_y$), leading to similar behavior in this regime.

\section{Conclusion}
\label{Conclusion}

In this work, we studied tunneling through a square barrier in a rotationally symmetric Rarita-Schwinger semi-metal. Because the low-energy spectrum contains two propagating sectors, $m=1/2$ and $m=3/2$, the barrier problem naturally involves multichannel scattering even at a fixed energy. We demonstrated that transmissions show a bias toward the spin projection $3/2$ channel. We also showed that when only one sector is injected, the transmission remains symmetric under $k_y\to -k_y$, while coherent mixed incidence of the two sectors leads to a clear directional asymmetry in the transmission probability.

The origin of this effect, we stress, is not an anisotropy of the dispersion or broken mirror symmetry. Rather, in the model considered here, the spectrum is rotationally symmetric, mirror symmetry is preserved, and the asymmetry instead arises from the phase structure of the multi-component wave functions. The two propagating sectors acquire different scattering and propagation phases, and for coherent mixed incidence, these phases enter interference terms in the transmitted amplitudes. As a result, the transmission can satisfy $T(k_y)\neq T(-k_y)$ even though both the Hamiltonian and the barrier are mirror symmetric.

This makes the effect noteworthy for two reasons. First, it shows that directional tunneling asymmetries can appear already in an idealized isotropic model, without invoking explicit band anisotropy. Second, it identifies the relevant mechanism as internal-state-dependent phase accumulation and coherent mode mixing. In this
sense, the effect is Imbert-Fedorov-like only at the level of
interface-induced phase differences between internal wave
components, rather than a claim of a conventional real-space
Imbert-Fedorov shift.

More broadly, our results suggest that tunneling in multiband systems
should be analyzed beyond band-structure symmetry alone. Another
natural extension is to consider junctions between Rarita--Schwinger
semimetals and normal or Dirac conductors, following analogous studies
of normal-metal--three-dimensional-Dirac-material junctions
\cite{SALEHI201564}. At such interfaces, the lead modes may couple to
superpositions of the $m=1/2$ and $m=3/2$ sectors, giving rise to
additional mode-conversion effects.

Moreover, in realistic materials, additional anisotropies may, of course, also contribute away from the node, but
asymmetric transport signatures should not automatically be attributed
to dispersion anisotropy. The present results therefore motivate a
reexamination of tunneling in other multichannel semimetals and related
systems to identify similar phase-driven directional effects.
\section{Acknowledgments}
O. Bouladiane acknowledges the support provided by CNRST in the framework of the program "PhD-Associate Scholarship -- PASS".
A. Al Luhaibi at KFUPM would like to acknowledge the support received under the University Funded Grant EC241024. M.V. gratefully acknowledges the support provided by the Deanship of Research Oversight and Coordination (DROC) and the Interdisciplinery Reasearch Center(IRC) for Advanced Quantum Computing (AQC) at King Fahd University of Petroleum \& Minerals (KFUPM) for funding his contribution to this work through research grant No. IN2607. 

The authors acknowledge use of ChatGPT 6.1 Sol (medium reasoning effort), ChatGPT 5.6 Sol, and Grammarly to assist with reorganizing parts of the manuscript, refining wording, and optimizing computational code, including translating existing Mathematica code into Python. Following their own checks, the authors also used ChatGPT to assist with additional consistency checks of the computations. These tools were used under the authors’ direction, and their outputs were reviewed by the authors. The authors take full responsibility for the scientific content and results presented in this work.

\bibliographystyle{unsrt}
\bibliography{literature}

@article{PhysRevD.5.787,
  title = {Calculation and Experimental Proof of the Transverse Shift Induced by Total Internal Reflection of a Circularly Polarized Light Beam},
  author = {Imbert, Christian},
  journal = {Phys. Rev. D},
  volume = {5},
  issue = {4},
  pages = {787--796},
  numpages = {0},
  year = {1972},
  month = {Feb},
  publisher = {American Physical Society},
  doi = {10.1103/PhysRevD.5.787},
  url = {https://link.aps.org/doi/10.1103/PhysRevD.5.787}
}

@article{Fedorov_2013,
doi = {10.1088/2040-8978/15/1/014002},
url = {https://doi.org/10.1088/2040-8978/15/1/014002},
year = {2013},
month = {jan},
publisher = {IOP Publishing},
volume = {15},
number = {1},
pages = {014002},
author = {Fedorov, Fedor I},
title = {To the theory of total reflection*},
journal = {Journal of Optics}
}

@article{PhysRevLett.93.083901,
  title = {Hall Effect of Light},
  author = {Onoda, Masaru and Murakami, Shuichi and Nagaosa, Naoto},
  journal = {Phys. Rev. Lett.},
  volume = {93},
  issue = {8},
  pages = {083901},
  numpages = {4},
  year = {2004},
  month = {Aug},
  publisher = {American Physical Society},
  doi = {10.1103/PhysRevLett.93.083901},
  url = {https://link.aps.org/doi/10.1103/PhysRevLett.93.083901}
}

@article{PhysRevLett.124.127602,
  title = {Interplay of Topology and Electron-Electron Interactions in Rarita-Schwinger-Weyl semimetals},
  author = {Boettcher, Igor},
  journal = {Phys. Rev. Lett.},
  volume = {124},
  issue = {12},
  pages = {127602},
  numpages = {6},
  year = {2020},
  month = {Mar},
  publisher = {American Physical Society},
  doi = {10.1103/PhysRevLett.124.127602},
  url = {https://link.aps.org/doi/10.1103/PhysRevLett.124.127602}
}

@article{PhysRevB.101.184503,
  title = {$d$-wave superconductivity and Bogoliubov-Fermi surfaces in Rarita-Schwinger-Weyl semimetals},
  author = {Link, Julia M. and Boettcher, Igor and Herbut, Igor F.},
  journal = {Phys. Rev. B},
  volume = {101},
  issue = {18},
  pages = {184503},
  numpages = {19},
  year = {2020},
  month = {May},
  publisher = {American Physical Society},
  doi = {10.1103/PhysRevB.101.184503},
  url = {https://link.aps.org/doi/10.1103/PhysRevB.101.184503}
}

@article{MANDAL2020126666,
title = {Transmission in pseudospin-1 and pseudospin-3/2 semimetals with linear dispersion through scalar and vector potential barriers},
journal = {Physics Letters A},
volume = {384},
number = {27},
pages = {126666},
year = {2020},
issn = {0375-9601},
doi = {https://doi.org/10.1016/j.physleta.2020.126666},
url = {https://www.sciencedirect.com/science/article/pii/S0375960120305338},
author = {Ipsita Mandal}
}

@Article{Mandal2023,
author={Mandal, Ipsita},
title={Transmission and conductance across junctions of isotropic and anisotropic three-dimensional semimetals},
journal={The European Physical Journal Plus},
year={2023},
month={Nov},
day={24},
volume={138},
number={11},
pages={1039},
issn={2190-5444},
doi={10.1140/epjp/s13360-023-04652-4},
url={https://doi.org/10.1140/epjp/s13360-023-04652-4}
}

@article{PhysRevB.99.241104,
  title = {Observation of multiple types of topological fermions in PdBiSe},
  author = {Lv, B. Q. and Feng, Z.-L. and Zhao, J.-Z. and Yuan, Noah F. Q. and Zong, A. and Luo, K. F. and Yu, R. and Huang, Y.-B. and Strocov, V. N. and Chikina, A. and Soluyanov, A. A. and Gedik, N. and Shi, Y.-G. and Qian, T. and Ding, H.},
  journal = {Phys. Rev. B},
  volume = {99},
  issue = {24},
  pages = {241104},
  numpages = {7},
  year = {2019},
  month = {Jun},
  publisher = {American Physical Society},
  doi = {10.1103/PhysRevB.99.241104},
  url = {https://link.aps.org/doi/10.1103/PhysRevB.99.241104}
}

@article{Bouhlal_2022,
   title={Tunneling Phase Diagrams in Anisotropic Multi‐Weyl Semimetals},
   volume={534},
   ISSN={1521-3889},
   url={http://dx.doi.org/10.1002/andp.202200267},
   DOI={10.1002/andp.202200267},
   number={11},
   journal={Annalen der Physik},
   publisher={Wiley},
   author={Bouhlal, Ahmed and Abbout, Adel and Jellal, Ahmed and Bahlouli, Hocine and Vogl, Michael},
   year={2022},
   month=Sept }

@article{BOUHLAL2021168563,
title = {Tunneling in an anisotropic cubic Dirac semi-metal},
journal = {Annals of Physics},
volume = {432},
pages = {168563},
year = {2021},
issn = {0003-4916},
doi = {https://doi.org/10.1016/j.aop.2021.168563},
url = {https://www.sciencedirect.com/science/article/pii/S000349162100169X},
author = {Ahmed Bouhlal and Ahmed Jellal and Hocine Bahlouli and Michael Vogl}
}

@article{PhysRevB.93.045113,
  title = {Semimetal with both Rarita-Schwinger-Weyl and Weyl excitations},
  author = {Liang, Long and Yu, Yue},
  journal = {Phys. Rev. B},
  volume = {93},
  issue = {4},
  pages = {045113},
  numpages = {6},
  year = {2016},
  month = {Jan},
  publisher = {American Physical Society},
  doi = {10.1103/PhysRevB.93.045113},
  url = {https://link.aps.org/doi/10.1103/PhysRevB.93.045113}
}

@article{Mandal_2024,
   title={Andreev bound states in superconductor-barrier-superconductor junctions of Rarita-Schwinger-Weyl semimetals},
   volume={503},
   ISSN={0375-9601},
   url={http://dx.doi.org/10.1016/j.physleta.2024.129410},
   DOI={10.1016/j.physleta.2024.129410},
   journal={Physics Letters A},
   publisher={Elsevier BV},
   author={Mandal, Ipsita},
   year={2024},
   month=Apr, pages={129410} }

@article{
doi:10.1126/science.aaf5037,
author = {Barry Bradlyn  and Jennifer Cano  and Zhijun Wang  and M. G. Vergniory  and C. Felser  and R. J. Cava  and B. Andrei Bernevig },
title = {Beyond Dirac and Weyl fermions: Unconventional quasiparticles in conventional crystals},
journal = {Science},
volume = {353},
number = {6299},
pages = {aaf5037},
year = {2016},
doi = {10.1126/science.aaf5037},
URL = {https://www.science.org/doi/abs/10.1126/science.aaf5037},
eprint = {https://www.science.org/doi/pdf/10.1126/science.aaf5037}}

@article{PhysRevLett.119.206402,
  title = {Multiple Types of Topological Fermions in Transition Metal Silicides},
  author = {Tang, Peizhe and Zhou, Quan and Zhang, Shou-Cheng},
  journal = {Phys. Rev. Lett.},
  volume = {119},
  issue = {20},
  pages = {206402},
  numpages = {6},
  year = {2017},
  month = {Nov},
  publisher = {American Physical Society},
  doi = {10.1103/PhysRevLett.119.206402},
  url = {https://link.aps.org/doi/10.1103/PhysRevLett.119.206402}
}

@article{PhysRevB.105.235403,
  title = {Circular dichroism as a probe for topology in three-dimensional semimetals},
  author = {Sekh, Sajid and Mandal, Ipsita},
  journal = {Phys. Rev. B},
  volume = {105},
  issue = {23},
  pages = {235403},
  numpages = {11},
  year = {2022},
  month = {Jun},
  publisher = {American Physical Society},
  doi = {10.1103/PhysRevB.105.235403},
  url = {https://link.aps.org/doi/10.1103/PhysRevB.105.235403}
}

@article{DOMBEY199941,
title = {Seventy years of the Klein paradox},
journal = {Physics Reports},
volume = {315},
number = {1},
pages = {41-58},
year = {1999},
issn = {0370-1573},
doi = {https://doi.org/10.1016/S0370-1573(99)00023-X},
url = {https://www.sciencedirect.com/science/article/pii/S037015739900023X},
author = {N. Dombey and A. Calogeracos}
}

@article{Pereira_2010,
doi = {10.1088/0268-1242/25/3/033002},
url = {https://doi.org/10.1088/0268-1242/25/3/033002},
year = {2010},
month = {feb},
publisher = {},
volume = {25},
number = {3},
pages = {033002},
author = {Pereira, J M and Peeters, F M and Chaves, A and Farias, G A},
title = {Klein tunneling in single and multiple barriers in graphene},
journal = {Semiconductor Science and Technology}
}

@article{PhysRevB.96.115448,
  title = {Imbert-Fedorov shift in Weyl semimetals: Dependence on monopole charge and intervalley scattering},
  author = {Wang, Luyang and Jian, Shao-Kai},
  journal = {Phys. Rev. B},
  volume = {96},
  issue = {11},
  pages = {115448},
  numpages = {7},
  year = {2017},
  month = {Sep},
  publisher = {American Physical Society},
  doi = {10.1103/PhysRevB.96.115448},
  url = {https://link.aps.org/doi/10.1103/PhysRevB.96.115448}
}

@article{RAZA2025130167,
title = {Goos-Hänchen shifts in transition metal dichalcogenides},
journal = {Physics Letters A},
volume = {531},
pages = {130167},
year = {2025},
issn = {0375-9601},
doi = {https://doi.org/10.1016/j.physleta.2024.130167},
url = {https://www.sciencedirect.com/science/article/pii/S0375960124008612},
author = {Mohsin Raza and Shamsher Ali and Muzamil Shah}
}

@article{RevModPhys.80.1337,
  title = {Colloquium: Andreev reflection and Klein tunneling in graphene},
  author = {Beenakker, C. W. J.},
  journal = {Rev. Mod. Phys.},
  volume = {80},
  issue = {4},
  pages = {1337--1354},
  numpages = {0},
  year = {2008},
  month = {Oct},
  publisher = {American Physical Society},
  doi = {10.1103/RevModPhys.80.1337},
  url = {https://link.aps.org/doi/10.1103/RevModPhys.80.1337}
}

@article{PhysRevLett.111.066803,
  title = {Chiral Tunneling in a Twisted Graphene Bilayer},
  author = {He, Wen-Yu and Chu, Zhao-Dong and He, Lin},
  journal = {Phys. Rev. Lett.},
  volume = {111},
  issue = {6},
  pages = {066803},
  numpages = {5},
  year = {2013},
  month = {Aug},
  publisher = {American Physical Society},
  doi = {10.1103/PhysRevLett.111.066803},
  url = {https://link.aps.org/doi/10.1103/PhysRevLett.111.066803}
}

@article{Culcer2020Transport2D,
  title   = {Transport in two-dimensional topological materials: recent developments in experiment and theory},
  author  = {Culcer, Dimitrie and Keser, Ayd{\i}n Cem and Li, Yongqing and Tkachov, Grigory},
  journal = {2D Materials},
  volume  = {7},
  number  = {2},
  pages   = {022007},
  year    = {2020},
  doi     = {10.1088/2053-1583/ab6ff7},
  publisher = {IOP Publishing},
  note    = {Topical Review, Open Access}
}

@article{Chang2018ChiralTopological,
  title   = {Topological quantum properties of chiral crystals},
  author  = {Chang, Guoqing and Wieder, Benjamin J. and Schindler, Frank and Sanchez, Daniel S. and Belopolski, Ilya and Huang, Shengyuan and Singh, Bahadur and Wu, Di and Chang, Tay-Rong and Neupert, Titus and Xu, Su-Yang and Hasan, M. Zahid},
  journal = {Nature Materials},
  volume  = {17},
  pages   = {978--985},
  year    = {2018},
  doi     = {10.1038/s41563-018-0169-3},
  publisher = {Nature Publishing Group}
}

@article{PhysRevB.68.205423,
  title = {Fabry-Perot interference and spin filtering in carbon nanotubes},
  author = {Pe\ifmmode \mbox{\c{c}}\else \c{c}\fi{}a, Claudia S. and Balents, Leon and Wiese, Kay J\"org},
  journal = {Phys. Rev. B},
  volume = {68},
  issue = {20},
  pages = {205423},
  numpages = {11},
  year = {2003},
  month = {Nov},
  publisher = {American Physical Society},
  doi = {10.1103/PhysRevB.68.205423},
  url = {https://link.aps.org/doi/10.1103/PhysRevB.68.205423}
}

@article{Dyakov2024ChiralLight,
  title   = {Chiral Light in Twisted Fabry--P{\'e}rot Cavities},
  author  = {Dyakov, Sergey A. and Salakhova, Natalia S. and Ignatov, Alexey V. and Fradkin, Ilia M. and Panov, Vitaly P. and Song, Jang-Kun and Gippius, Nikolay A.},
  journal = {Advanced Optical Materials},
  volume  = {12},
  pages   = {2302502},
  year    = {2024},
  doi     = {10.1002/adom.202302502},
  publisher = {Wiley}
}

@article{doi:10.1126/science.1102896,
author = {K. S. Novoselov  and A. K. Geim  and S. V. Morozov  and D. Jiang  and Y. Zhang  and S. V. Dubonos  and I. V. Grigorieva  and A. A. Firsov },
title = {Electric Field Effect in Atomically Thin Carbon Films},
journal = {Science},
volume = {306},
number = {5696},
pages = {666-669},
year = {2004},
doi = {10.1126/science.1102896},
URL = {https://www.science.org/doi/abs/10.1126/science.1102896},
eprint = {https://www.science.org/doi/pdf/10.1126/science.1102896}}

@article{Geim2007,
  title={The rise of graphene},
  author={Geim, A. K. and Novoselov, K. S.},
  journal={Nature Materials},
  volume={6},
  pages={183--191},
  year={2007}
}

@article{Katsnelson2006,
  title={Chiral tunnelling and the Klein paradox in graphene},
  author={Katsnelson, M. I. and Novoselov, K. S. and Geim, A. K.},
  journal={Nature Physics},
  volume={2},
  pages={620--625},
  year={2006}
}

@article{RevModPhys.81.109,
  title = {The electronic properties of graphene},
  author = {Castro Neto, A. H. and Guinea, F. and Peres, N. M. R. and Novoselov, K. S. and Geim, A. K.},
  journal = {Rev. Mod. Phys.},
  volume = {81},
  issue = {1},
  pages = {109--162},
  numpages = {0},
  year = {2009},
  month = {Jan},
  publisher = {American Physical Society},
  doi = {10.1103/RevModPhys.81.109},
  url = {https://link.aps.org/doi/10.1103/RevModPhys.81.109}
}

@article{Young2009,
  author    = {Young, Andrea F. and Kim, Philip},
  title     = {Quantum interference and Klein tunnelling in graphene heterojunctions},
  journal   = {Nature Physics},
  volume    = {5},
  number    = {3},
  pages     = {222--226},
  year      = {2009},
  doi       = {10.1038/nphys1198},
  publisher = {Nature Publishing Group}
}

@article{Katsnelson2006ZB,
  title={Zitterbewegung, chirality, and minimal conductivity in graphene},
  author={Katsnelson, M. I.},
  journal={European Physical Journal B},
  volume={51},
  pages={157--160},
  year={2006}
}

@article{PhysRevLett.102.146804,
  title = {Quantum Goos-H\"anchen Effect in Graphene},
  author = {Beenakker, C. W. J. and Sepkhanov, R. A. and Akhmerov, A. R. and Tworzyd\l{}o, J.},
  journal = {Phys. Rev. Lett.},
  volume = {102},
  issue = {14},
  pages = {146804},
  numpages = {4},
  year = {2009},
  month = {Apr},
  publisher = {American Physical Society},
  doi = {10.1103/PhysRevLett.102.146804},
  url = {https://link.aps.org/doi/10.1103/PhysRevLett.102.146804}
}

@article{PhysRevB.62.10696,
  title = {Theory of light propagation in strongly modulated photonic crystals: Refractionlike behavior in the vicinity of the photonic band gap},
  author = {Notomi, M.},
  journal = {Phys. Rev. B},
  volume = {62},
  issue = {16},
  pages = {10696--10705},
  numpages = {0},
  year = {2000},
  month = {Oct},
  publisher = {American Physical Society},
  doi = {10.1103/PhysRevB.62.10696},
  url = {https://link.aps.org/doi/10.1103/PhysRevB.62.10696}
}

@article{PhysRevLett.96.073903,
  title = {Conservation of Angular Momentum, Transverse Shift, and Spin Hall Effect in Reflection and Refraction of an Electromagnetic Wave Packet},
  author = {Bliokh, Konstantin Yu. and Bliokh, Yury P.},
  journal = {Phys. Rev. Lett.},
  volume = {96},
  issue = {7},
  pages = {073903},
  numpages = {4},
  year = {2006},
  month = {Feb},
  publisher = {American Physical Society},
  doi = {10.1103/PhysRevLett.96.073903},
  url = {https://link.aps.org/doi/10.1103/PhysRevLett.96.073903}
}

@article{PhysRevB.92.081201,
  title = {Topological nodal line semimetals with and without spin-orbital coupling},
  author = {Fang, Chen and Chen, Yige and Kee, Hae-Young and Fu, Liang},
  journal = {Phys. Rev. B},
  volume = {92},
  issue = {8},
  pages = {081201},
  numpages = {5},
  year = {2015},
  month = {Aug},
  publisher = {American Physical Society},
  doi = {10.1103/PhysRevB.92.081201},
  url = {https://link.aps.org/doi/10.1103/PhysRevB.92.081201}
}

@article{PhysRevResearch.2.013088,
  title = {Thermal and gravitational chiral anomaly induced magneto-transport in Weyl semimetals},
  author = {Das, Kamal and Agarwal, Amit},
  journal = {Phys. Rev. Res.},
  volume = {2},
  issue = {1},
  pages = {013088},
  numpages = {9},
  year = {2020},
  month = {Jan},
  publisher = {American Physical Society},
  doi = {10.1103/PhysRevResearch.2.013088},
  url = {https://link.aps.org/doi/10.1103/PhysRevResearch.2.013088}
}

@article{PhysRevB.89.075124,
  title = {Anomalous transport of Weyl fermions in Weyl semimetals},
  author = {Landsteiner, Karl},
  journal = {Phys. Rev. B},
  volume = {89},
  issue = {7},
  pages = {075124},
  numpages = {11},
  year = {2014},
  month = {Feb},
  publisher = {American Physical Society},
  doi = {10.1103/PhysRevB.89.075124},
  url = {https://link.aps.org/doi/10.1103/PhysRevB.89.075124}
}

@article{RevModPhys.90.015001,
  title = {Weyl and Dirac semimetals in three-dimensional solids},
  author = {Armitage, N. P. and Mele, E. J. and Vishwanath, Ashvin},
  journal = {Rev. Mod. Phys.},
  volume = {90},
  issue = {1},
  pages = {015001},
  numpages = {57},
  year = {2018},
  month = {Jan},
  publisher = {American Physical Society},
  doi = {10.1103/RevModPhys.90.015001},
  url = {https://link.aps.org/doi/10.1103/RevModPhys.90.015001}
}

@article{PhysRevLett.115.156602,
  title = {Topological Imbert-Fedorov Shift in Weyl Semimetals},
  author = {Jiang, Qing-Dong and Jiang, Hua and Liu, Haiwen and Sun, Qing-Feng and Xie, X. C.},
  journal = {Phys. Rev. Lett.},
  volume = {115},
  issue = {15},
  pages = {156602},
  numpages = {5},
  year = {2015},
  month = {Oct},
  publisher = {American Physical Society},
  doi = {10.1103/PhysRevLett.115.156602},
  url = {https://link.aps.org/doi/10.1103/PhysRevLett.115.156602}
}

@article{ZHU2026170301,
title = {Transport properties of the pseudospin-3/2 Dirac–Weyl fermions in the double-barrier-modulated two-dimensional system},
journal = {Annals of Physics},
volume = {485},
pages = {170301},
year = {2026},
issn = {0003-4916},
doi = {https://doi.org/10.1016/j.aop.2025.170301},
url = {https://www.sciencedirect.com/science/article/pii/S0003491625003835},
author = {Rui Zhu}
}

@article{Gupta2001SpinCoherence,
author = {J. A. Gupta  and R. Knobel  and N. Samarth  and D. D. Awschalom },
title = {Ultrafast Manipulation of Electron Spin Coherence},
journal = {Science},
volume = {292},
number = {5526},
pages = {2458-2461},
year = {2001},
doi = {10.1126/science.1061169},
URL = {https://www.science.org/doi/abs/10.1126/science.1061169},
eprint = {https://www.science.org/doi/pdf/10.1126/science.1061169}}

@Article{Fiederling1999SpinInjection,
author={Fiederling, R.
and Keim, M.
and Reuscher, G.
and Ossau, W.
and Schmidt, G.
and Waag, A.
and Molenkamp, L. W.},
title={Injection and detection of a spin-polarized current in a light-emitting diode},
journal={Nature},
year={1999},
month={Dec},
day={01},
volume={402},
number={6763},
pages={787-790},
issn={1476-4687},
doi={10.1038/45502},
url={https://doi.org/10.1038/45502}
}

@Article{Ohno1999SpinInjection,
author={Ohno, Y.
and Young, D. K.
and Beschoten, B.
and Matsukura, F.
and Ohno, H.
and Awschalom, D. D.},
title={Electrical spin injection in a ferromagnetic semiconductor heterostructure},
journal={Nature},
year={1999},
month={Dec},
day={01},
volume={402},
number={6763},
pages={790-792},
issn={1476-4687},
doi={10.1038/45509},
url={https://doi.org/10.1038/45509}
}

@article{SanchezMartinez2019Optical,
  title = {Linear optical conductivity of chiral multifold fermions},
  author = {S{\'a}nchez-Mart{\'{\i}}nez, Miguel-{\'A}ngel and de Juan, Fernando and Grushin, Adolfo G.},
  journal = {Phys. Rev. B},
  volume = {99},
  issue = {15},
  pages = {155145},
  numpages = {14},
  year = {2019},
  month = {Apr},
  publisher = {American Physical Society},
  doi = {10.1103/PhysRevB.99.155145},
  url = {https://link.aps.org/doi/10.1103/PhysRevB.99.155145}
}

@article{PhysRevB.31.6207,
  title = {Generalized many-channel conductance formula with application to small rings},
  author = {B{\"u}ttiker, M. and Imry, Y. and Landauer, R. and Pinhas, S.},
  journal = {Phys. Rev. B},
  volume = {31},
  issue = {10},
  pages = {6207--6215},
  numpages = {0},
  year = {1985},
  month = {May},
  publisher = {American Physical Society},
  doi = {10.1103/PhysRevB.31.6207},
  url = {https://link.aps.org/doi/10.1103/PhysRevB.31.6207}
}

@article{PhysRevB.40.1456,
  title = {Landauer's conductance formula and its generalization to finite voltages},
  author = {Bagwell, Philip F. and Orlando, Terry P.},
  journal = {Phys. Rev. B},
  volume = {40},
  issue = {3},
  pages = {1456--1464},
  numpages = {0},
  year = {1989},
  month = {Jul},
  publisher = {American Physical Society},
  doi = {10.1103/PhysRevB.40.1456},
  url = {https://link.aps.org/doi/10.1103/PhysRevB.40.1456}
}

@article{PhysRevB.110.205406,
  title = {Tunneling valley Hall effect induced by coherent geometric phase},
  author = {Zeng, W.},
  journal = {Phys. Rev. B},
  volume = {110},
  issue = {20},
  pages = {205406},
  numpages = {9},
  year = {2024},
  month = {Nov},
  publisher = {American Physical Society},
  doi = {10.1103/PhysRevB.110.205406},
  url = {https://link.aps.org/doi/10.1103/PhysRevB.110.205406}
}

@article{SALEHI201564,
title = {Quantum transport through 3D Dirac materials},
journal = {Annals of Physics},
volume = {359},
pages = {64-72},
year = {2015},
issn = {0003-4916},
doi = {https://doi.org/10.1016/j.aop.2015.04.007},
url = {https://www.sciencedirect.com/science/article/pii/S0003491615001402},
author = {M. Salehi and S.A. Jafari}
}

\newpage
\appendix

\section{Details of mode-matching formulation}
\label{appendix}



In this appendix we discuss tunneling properties of the Rarita-Schwinger semi-metal in more technical detail.\\
First we recall that in each region, the wave function can be written as a superposition of right- and left-moving eigenspinors. It is convenient to collect the spinors into matrices and the corresponding amplitudes into coefficient vectors.
The general solutions in the three regions are
\begin{align}
\Psi_{\mathrm I}(x,y) &= G_1\,M_1(x)\,C_1\,e^{ik_y y},
\\
\Psi_{\mathrm {II}}(x,y) &= H_2\,M_2(x)\,C_2\,e^{ik_y y},
\\
\Psi_{\mathrm {III}}(x,y) &= G_1\,M_1(x)\,C_3\,e^{ik_y y}.
\end{align}
Here, the vectors $C_1$, $C_2$, and $C_3$ contain the amplitudes of the propagating modes in regions I, II, and III, respectively.

For the outer regions, the spinor matrix is
\begin{align}
G_1=
\begin{pmatrix}
- e^{-i3\phi_{1/2}}  &  e^{i3\phi_{1/2}} & e^{-i3\phi_{3/2}} & - e^{i3\phi_{3/2}} \\
-\frac{1}{\sqrt{3}}e^{-i2\phi_{1/2}} & -\frac{1}{\sqrt{3}}e^{i2\phi_{1/2}}
& \sqrt{3}e^{-i2\phi_{3/2}} & \sqrt{3}e^{i2\phi_{3/2}} \\
\frac{1}{\sqrt{3}}e^{-i\phi_{1/2}} & -\frac{1}{\sqrt{3}}e^{i\phi_{1/2}}
& \sqrt{3}e^{-i\phi_{3/2}} & -\sqrt{3}e^{i\phi_{3/2}} \\
1 & 1 & 1 & 1
\end{pmatrix},
\end{align}
and the corresponding diagonal propagation matrix is
\begin{align}
M_1(x)=
\begin{pmatrix}
e^{ik_{1/2}x} & 0 & 0 & 0 \\
0 & e^{-ik_{1/2}x} & 0 & 0 \\
0 & 0 & e^{ik_{3/2}x} & 0 \\
0 & 0 & 0 & e^{-ik_{3/2}x}
\end{pmatrix}.
\end{align}
The coefficient vectors in regions I and III are
\begin{align}
C_1=
\begin{pmatrix}
\delta_{1/2} \\
r_{1/2} \\
\delta_{3/2} \\
r_{3/2}
\end{pmatrix},
\qquad
C_3=
\begin{pmatrix}
t_{1/2} \\
0 \\
t_{3/2} \\
0
\end{pmatrix}.
\end{align}
Inside the barrier, the corresponding spinor matrix is
\begin{align}
H_2=
\begin{pmatrix}
- e^{-i3\theta_{1/2}}  &  e^{i3\theta_{1/2}} & e^{-i3\theta_{3/2}} & - e^{i3\theta_{3/2}} \\
-\frac{1}{\sqrt{3}}e^{-i2\theta_{1/2}} & -\frac{1}{\sqrt{3}}e^{i2\theta_{1/2}}
& \sqrt{3}e^{-i2\theta_{3/2}} & \sqrt{3}e^{i2\theta_{3/2}} \\
\frac{1}{\sqrt{3}}e^{-i\theta_{1/2}} & -\frac{1}{\sqrt{3}}e^{i\theta_{1/2}}
& \sqrt{3}e^{-i\theta_{3/2}} & -\sqrt{3}e^{i\theta_{3/2}} \\
1 & 1 & 1 & 1
\end{pmatrix},
\end{align}
with the propagation matrix
\begin{align}
M_2(x)=
\begin{pmatrix}
e^{iq_{1/2}x} & 0 & 0 & 0 \\
0 & e^{-iq_{1/2}x} & 0 & 0 \\
0 & 0 & e^{iq_{3/2}x} & 0 \\
0 & 0 & 0 & e^{-iq_{3/2}x}
\end{pmatrix},
\end{align}
and the coefficient vector
\begin{align}
C_2=
\begin{pmatrix}
a_{1/2}^{+} \\
a_{1/2}^{-} \\
a_{3/2}^{+} \\
a_{3/2}^{-}
\end{pmatrix}.
\end{align}


The unknown amplitudes $r_{1/2}$, $r_{3/2}$, $a_{1/2}^{\pm}$, $a_{3/2}^{\pm}$, $t_{1/2}$, and $t_{3/2}$ are determined by continuity of the wave function at the two interfaces $x=0$ and $x=L$:
\begin{align}
&\Psi_{\mathrm I}(0,y) = \Psi_{\mathrm {II}}(0,y),
\\
&\Psi_{\mathrm {II}}(L,y) = \Psi_{\mathrm {III}}(L,y).
\end{align}
In matrix form, these conditions become
\begin{align}
&G_1\,M_1(0)\,C_1 = H_2\,M_2(0)\,C_2,
\\
&H_2\,M_2(L)\,C_2 = G_1\,M_1(L)\,C_3.
\end{align}
Eliminating $C_2$ gives a direct relation between the coefficients in regions I and III:
\begin{equation}
C_1=\mathcal N\,C_3,
\end{equation}
where the transfer matrix $\mathcal N$ is
\begin{equation}
\mathcal N
=
M_1^{-1}(0)\,G_1^{-1}\,H_2\,M_2(0)\,M_2^{-1}(L)\,H_2^{-1}\,G_1\,M_1(L).
\end{equation}
Writing this relation explicitly,
\begin{align}
\begin{pmatrix}
\delta_{1/2} \\
r_{1/2} \\
\delta_{3/2} \\
r_{3/2}
\end{pmatrix}
=
\begin{pmatrix}
\mathcal N_{11} & \mathcal N_{12} & \mathcal N_{13} & \mathcal N_{14} \\
\mathcal N_{21} & \mathcal N_{22} & \mathcal N_{23} & \mathcal N_{24} \\
\mathcal N_{31} & \mathcal N_{32} & \mathcal N_{33} & \mathcal N_{34} \\
\mathcal N_{41} & \mathcal N_{42} & \mathcal N_{43} & \mathcal N_{44}
\end{pmatrix}
\begin{pmatrix}
t_{1/2} \\
0 \\
t_{3/2} \\
0
\end{pmatrix}.
\label{sool}
\end{align}
From this, the transmission amplitudes are obtained as
\begin{align}
t_{3/2} &=
\frac{\mathcal N_{11}\delta_{3/2}-\mathcal N_{31}\delta_{1/2}}
{\mathcal N_{11}\mathcal N_{33}-\mathcal N_{13}\mathcal N_{31}},
\quad
t_{1/2} =
\frac{\mathcal N_{33}\delta_{1/2}-\mathcal N_{13}\delta_{3/2}}
{\mathcal N_{11}\mathcal N_{33}-\mathcal N_{13}\mathcal N_{31}}.
\end{align}

The reflection amplitudes then follow from
\begin{align}
r_{1/2} &= \mathcal N_{21}t_{1/2}+\mathcal N_{23}t_{3/2},
\quad
r_{3/2} &= \mathcal N_{41}t_{1/2}+\mathcal N_{43}t_{3/2}.
\end{align}
\end{document}